\documentclass[graybox]{svmult}

\usepackage{url}       
\usepackage{mathptmx}       
\usepackage{helvet}         
\usepackage{courier}        
\usepackage{makeidx}         
\usepackage{graphicx}        
\usepackage{multicol}        
\usepackage[bottom]{footmisc}
\usepackage[T1]{fontenc}
\usepackage{amssymb,amsmath}
\usepackage{ifxetex,ifluatex}
\usepackage{fixltx2e} 
\IfFileExists{upquote.sty}{\usepackage{upquote}}{}
\ifnum 0\ifxetex 1\fi\ifluatex 1\fi=0 
\else 
  \ifxetex
    \usepackage{mathspec}
    \usepackage{xltxtra,xunicode}
  \else
    \usepackage{fontspec}
  \fi
  \defaultfontfeatures{Mapping=tex-text,Scale=MatchLowercase}
  
\fi
\IfFileExists{microtype.sty}{\usepackage{microtype}}{}
\usepackage{color}
\usepackage{fancyvrb}

\DefineVerbatimEnvironment{Highlighting}{Verbatim}{commandchars=\\\{\},fontsize=\small}
\usepackage{mdframed}
\newenvironment{Shaded}{\begin{mdframed}[leftline=false, rightline=false]}{\end{mdframed}}
\newcommand{\KeywordTok}[1]{\textcolor[rgb]{0.00,0.44,0.13}{\textbf{{#1}}}}
\newcommand{\DataTypeTok}[1]{\textcolor[rgb]{0.56,0.13,0.00}{{#1}}}
\newcommand{\DecValTok}[1]{\textcolor[rgb]{0.25,0.63,0.44}{{#1}}}

\newcommand{\FloatTok}[1]{\textcolor[rgb]{0.25,0.63,0.44}{{#1}}}

\newcommand{\StringTok}[1]{\textcolor[rgb]{0.25,0.44,0.63}{{#1}}}
\newcommand{\CommentTok}[1]{\textcolor[rgb]{0.38,0.63,0.69}{\textit{{#1}}}}
\newcommand{\OtherTok}[1]{\textcolor[rgb]{0.00,0.44,0.13}{{#1}}}

\newcommand{\NormalTok}[1]{{#1}}
\usepackage{graphicx}
\makeatletter
\def\ScaleIfNeeded{%
  \ifdim\Gin@nat@width>\linewidth
    \linewidth
  \else
    \Gin@nat@width
  \fi
}
\makeatother
\let\Oldincludegraphics\includegraphics
{%
 \catcode`\@=11\relax%
 \gdef\includegraphics{\@ifnextchar[{\Oldincludegraphics}{\Oldincludegraphics[width=\ScaleIfNeeded]}}%
}%
\ifxetex
  \usepackage[setpagesize=false, 
              unicode=false, 
              xetex]{hyperref}
\else
  \usepackage[unicode=true]{hyperref}
\fi
\hypersetup{breaklinks=true,
            bookmarks=true,
            pdfauthor={},
            pdftitle={R Notebook},
            colorlinks=true,
            citecolor=blue,
            urlcolor=blue,
            linkcolor=magenta,
            pdfborder={0 0 0}}
 \DeclareTextCommandDefault{\nobreakspace}{\leavevmode\nobreak\ } 

\makeindex             
\usepackage[T1]{fontenc}
\usepackage[numbered,framed]{matlab-prettifier}

\lstMakeShortInline"

\begin{document}                       

\title*{Seeing the wood for the trees: a forest of methods for optimisation and omic-network integration in metabolic modelling}
\titlerunning{Optimisation methods in metabolic modelling} 
\author{Supreeta Vijayakumar*, Max Conway*, Pietro Li\'o and Claudio Angione}


\institute{Supreeta Vijayakumar \at Department of Computer Science and Information Systems, Teesside University, UK\\
Supreeta Vijayakumar is a PhD student at the Department of Computer Science and Information Systems, Teesside University, UK. Her research focuses on the integration of multi-omic data and machine learning algorithms for constraint-based modelling of microbial communities.\\ 
\email{s.vijayakumar@tees.ac.uk}
\and Max Conway \at Computer Laboratory, University of Cambridge, UK\\
Max Conway is a PhD student at the Computer Laboratory, University of Cambridge, UK. His research interests include machine learning to extract structure from metabolic and multi-omic models.\\
\email{max.conway@cl.cam.ac.uk}
\and Pietro Li\'o \at Computer Laboratory, University of Cambridge, UK\\
Pietro Li\'o is a Reader in Computational Biology at the Computer Laboratory, University of Cambridge. His affiliations also include the Cambridge Computational Biology Institute. His work spans Machine Learning and computational models for health Big Data, personalised medicine research, multi-scale/multi-omic/multi-physics modelling and data integration methods.\\
\email{pietro.lio@cl.cam.ac.uk}
\and Claudio Angione \at Department of Computer Science and Information Systems, Teesside University, UK\\
Claudio Angione is a Senior Lecturer in Computer Science at the Department of Computer Science and Information Systems, Teesside University, UK. He holds a PhD in Computer Science from the University of Cambridge, UK. His research interests include cancer metabolism, machine learning, systems biology and optimisation of genome-scale and poly-omic models.\\
\email{c.angione@tees.ac.uk}
\and *These authors contributed equally to this work}

%
%
\maketitle

\vspace{-2cm}
\begin{itemize}
\item Being downstream of gene expression, metabolism is being increasingly used as an indicator of the phenotypic outcome for drugs and therapies.
\item We present an online resource and up-to-date classification of existing methods and tools for poly-omic modelling of metabolism in systems biology.
\item We provide a hands-on tutorial for multi-objective optimisation of metabolic models in R.
\item We finally discuss the implementation of multi-view machine-learning approaches in poly-omic data integration.
\end{itemize}

This is a pre-copyedited, author-produced version of an article accepted for publication in Briefings in Bioinformatics following peer review. The version of record ``S. Vijayakumar, M. Conway, P. Li\'o, and C. Angione, ``Seeing the wood for the trees: a forest of methods for optimisation and omic-network integration in metabolic modelling'', Briefings in Bioinformatics, bbx053, 2017'' is available online at: \url{https://academic.oup.com/bib/advance-article-abstract/doi/10.1093/bib/bbx053/3858337}.

\newpage\abstract{Metabolic modelling  has entered a mature phase with dozens of methods and software implementations available to the practitioner and the theoretician. It is not easy for a modeller to be able to see the wood (or the forest) for the trees. Driven by this analogy, we here present a ``forest'' of principal methods used for constraint-based modelling in systems biology. This provides a tree-based view of methods available to prospective modellers, also available in interactive version at \url{http://modellingmetabolism.net}, where it will be kept updated with new methods after the publication of the present manuscript. Our updated classification of existing methods and tools highlights the most promising in the different branches, with the aim to develop a vision of how existing methods could hybridise and become more complex. We then provide the first hands-on tutorial for multi-objective optimisation of metabolic models in R. We finally discuss the implementation of multi-view machine-learning approaches in poly-omic integration. Throughout this work, we demonstrate the optimisation of trade-offs between multiple metabolic objectives, with a focus on omic data integration through machine learning. We anticipate that the combination of a survey, a perspective on multi-view machine learning, and a step-by-step R tutorial should be of interest for both the beginner and the advanced user.}

\section{Introduction}
\label{sec:intro}

Metabolism is the indispensable set of biochemical reactions in a cell that maintain its living state. Constraint-based reconstruction and analysis (COBRA) are a group of techniques that are commonly used for the mathematical and computational modelling of metabolic networks at the whole-genome scale. Genome-scale metabolic models are available in online repositories such as the Kyoto Encyclopedia of Genes and Genomes (KEGG) {\cite{kanehisa2000kegg}}, the Biochemical Genetic and Genomic (BiGG) knowledge-base {\cite{schellenberger2010bigg}}, the BioCyc collection of pathway/genome databases {\cite{karp2005expansion}}, MetaNetX \cite{moretti2016metanetx} and the ModelSEED database {\cite{henry2010high}}. Principally, the preparation of a genome-scale metabolic model involves the reconstruction of all metabolic reactions taking place in the organism supplemented with functional annotation of genes, metabolites and pathways. Depending on the quality of the reconstruction, processes of manual curation and gap-filling may also be required \cite{prigent2017meneco}.  Predictions obtained from genome-scale metabolic models can be reconciled with \textit{in-vivo} findings and used to identify current gaps in our knowledge of metabolism {\cite{mienda2016genome}}.

There are often inconsistencies between models and experimental data, such as when an outcome is falsely predicted by the model (false positive) or when an experimentally observed outcome is not predicted (false negative). Algorithms such as Grow Match {\cite{kumar2009growmatch}}, SMILEY {\cite{reed2006systems}} and optimal metabolic network identification (OMNI) {\cite{herrgaard2006identification}} correct for such inconsistencies by suggesting adjustments for improving model accuracy. Reducing disparity between predicted and experimentally measured fluxes presents opportunities to devise new strategies for biological discovery {\cite{o2015using}}. GapFind and GapFill are jointly designed optimisation procedures to identify `problem' metabolites which cannot be produced or consumed in the network and then propose mechanisms to restore pathway connectivity for these metabolites {\cite{kumar2007optimization}}. fastGapFill is an extension of FASTCORE which incorporates flux and stoichiometric consistency into the gap-filling process {\cite{thiele2014fastgapfill}}. Metabolic reconstruction via functional genomics (MIRAGE) conducts gap-filling by integrating with functional genomics data to estimate the probability of including each reaction from a universal database of gap-filling reactions in the reconstructed network {\cite{vitkin2012mirage}}. This enables selection of the set of reactions whose addition is most likely to result in a fully functional model when flux analysis is repeated. Many models also integrate signalling and regulatory pathways with metabolic networks in order to add information regarding underlying mechanisms, consequently improving flux predictions {\cite{palsson2015systems}}.

Here, we present the foundations of constraint-based metabolic modelling as well as recent advances, in the form of a `forest' of analytical tools and methods comprising algorithms and their software implementations. As such techniques are likely to expand and diversify over time, this schematic is also available in interactive version at \url{http://modellingmetabolism.net}, where it will be updated as newer methods are developed. We believe that classifying existing methods by their purpose or mode of implementation and defining their strengths and limitations will greatly facilitate the selection of methods for prospective modellers. In this regard, authors of new tools and methods are invited to contact us in order to include these in the interactive version of our figure.

In the following sections, we describe the main approaches currently used for constraint-based metabolic modelling, with the inclusion of many recent developments which we consider to be significant. These methods are divided into unbiased and biased approaches, the latter of which includes (i) a comprehensive review of flux balance analysis (FBA) and its specific variants, which apply different types of constraints for the prediction of metabolic fluxes (ii) regulatory methods, for which constraints are derived from external sources for designing context- or condition-specific metabolic models, and (iii) methods for the simulation of genetic perturbations and selection of the objective function. Following this, a detailed discussion of methods for performing multi-objective optimisation forms the basis of our tutorial for genetic design by multi-objective optimisation (GDMO) in R. Finally, we include a perspective that evaluates the potential use of multi-view machine learning techniques for the analysis of multi-omic metabolic models, which we regard as an important venture for the future of metabolic modelling.

\section{Methods for constraint-based metabolic modelling}
\label{sec:2}

\begin{figure}
\centering
\includegraphics[width=\columnwidth]{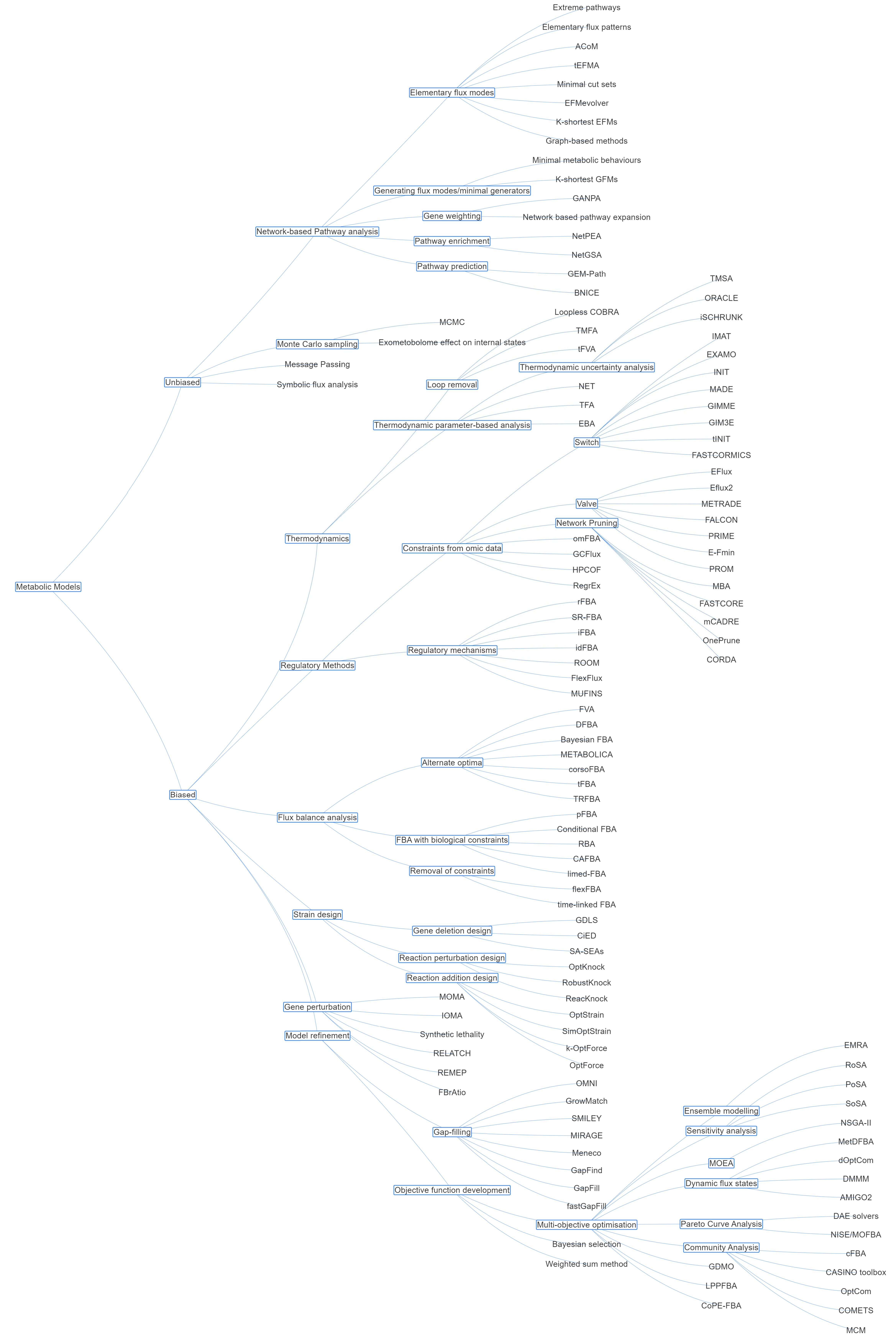} 
\caption{A forest of methods based on constraint-based reconstruction and analysis (COBRA). Network-based pathway analysis describes the simplest configurations of metabolic pathways at steady state. Monte Carlo methods allow for uniform sampling of the solution space to compute the flux as a probability distribution for each pathway in the network. Thermodynamically infeasible fluxes may be eliminated using loop removal or thermodynamic parameter-based analysis. Model refinement may be carried out through the selection of single or multiple objectives or gap-filling techniques. Gene perturbation helps to establish the essentiality of genes and the most efficient pathways for the production of specific metabolites. Following this, strain design for metabolic engineering may involve strategies for gene deletion, reaction perturbation, or reaction addition. Alternate optimal solutions may be yielded by variations of flux balance analysis. Other variants may specify the addition of specific constraints or their removal. Additional constraints may be introduced through the inclusion of multi-omic data or gene regulatory mechanisms. Note that some methods may belong to multiple categories, but for clarity we classify each method according to its main contribution. An interactive version of this figure is maintained and updated at \url{http://modellingmetabolism.net}.}
\label{fig:network}
\end{figure}

Figure \ref{fig:network} depicts the `forest' of methods commonly used for constraint-based modelling of metabolic networks, following and updating the framework proposed by Lewis et al.~{\cite{lewis2012constraining}}. Methodological approaches are broadly divided into biased and unbiased; the former necessitates the definition of an objective function by the network, whereas the latter relies on determining a subset of statistically analysable functional states whilst searching the entirety of the solution space {\cite{lee2009systems}}. 

\subsection{Unbiased methods}
\label{sec:3}

Unbiased methods search the entirety of the solution space and find a subset of statistically analysable functional states without requiring the definition of an objective function.
Network-based pathway analysis comprises a large family of unbiased methods assessing the main properties of biochemical pathways {\cite{papin2004comparison}}. Gene Association Network-based Pathway Analysis (GANPA) improves upon this process by adding gene weighting to determine gene non-equivalence within pathways {\cite{fang2012network}}. Similarly, a novel method was recently proposed for assessing the significance of pathways by constructing weighted gene-gene interaction networks for normal and cancerous tissue samples {\cite{zhang2016network}}. These interaction networks were subsequently used to expand pathways for each set of samples and compare their topologies. Approaches based on network-based pathway enrichment analysis aim to identify a greater number of gene interactions. For example, NetPEA utilises a protein-protein interaction (PPI) network combined with random walk to include information from high-throughput networks as well as known pathways {\cite{liu2013network}}. A combination of network estimation with condition-specific omic data has been used to refine the NetGSA framework, thereby improving the ability to detect differential activity in pathways {\cite{ma2016network}}. 

Using network-based pathway analysis, different methods may be used to calculate a set of routes through the reaction network and the corresponding kernel matrices which represent their stoichiometry. Elementary flux modes (EFMs) describe the minimal, non-decomposable set of pathways operating within a steady-state system; these are found by solving the steady-state condition following the iterative removal of single reactions until a valid flux distribution can no longer be calculated {\cite{zanghellini2013elementary}}. As this process often yields a combinatorial explosion of common functional motifs, a variation of the Agglomeration of Common Motifs (ACoM) method can be used to cluster these motifs, allowing for an overlap between classes {\cite{peres2011acom}}. Alternatively, a single EFM may be determined by solving an optimisation problem using EFMevolver {\cite{kaleta2009efmevolver}}, which can draw attention to significant EFMs. A method known as \textit{K}-shortest EFMs enumerates EFMs in order of their number of reactions and has been applied to genome-scale networks {\cite{de2009computing}}; the shortest pathways are of interest as they typically carry the highest flux and are easily manipulable. 

A minimal generating set is the smallest set necessary to define the geometry of the flux space using a null-space algorithm, the elements of which are known as generating flux modes (GFMs) or minimal generators {\cite{wagner2005geometry}}. A variant of this method can be used to find specific subsets of GFMs in a process which does not result in a combinatorial explosion, thus making them easier to compute {\cite{rezola2011exploring}}. Methods using minimal descriptions of the flux cone, such as minimal generators {\cite{urbanczik2005improved}} and minimal metabolic behaviours {\cite{larhlimi2009new}} aim to reduce the dimensionality of the flux cone. A recent method prioritises the search for the shortest path between a pair of end nodes based on graph theory {\cite{hidalgo2016new}}. However, the validity of this approach has been questioned as reaction stoichiometry is overlooked {\cite{rezola2014advances}}. tEFMA {\cite{gerstl2015metabolomics}} removes thermodynamically-infeasible EFMs using network-embedded thermodynamic (NET) analysis {\cite{kummel2006putative}}. The incorporation of thermodynamic constraints helps to select for physiologically significant EFMs, which become more difficult to detect as the size and complexity of networks increases. Identifying the largest thermodynamically consistent sets (LTCSs) in these EFMs can further characterise condition-specific metabolic capabilities in the thermodynamically-feasible regions of the flux cone {\cite{gerstl2016sets}}. 

Extreme pathways can be described as being a systemically independent subset of EFMs {\cite{papin2004comparison}}. They are characterised by a set of convex basis vectors used to represent the edges of the steady-state solution space and consist of the minimum number of reactions needed to exist as a functional unit {\cite{schilling2000theory}}. As opposed to many of methods described previously, extreme currents aim to increase dimensionality by describing non-decomposable EFMs situated both within and on the boundaries of the flux regions; these arise as a result of partitioning each reversible reaction into two irreversible reactions {\cite{clarke1988stoichiometric}}. Minimal cut sets (MCSs) are another variant of EFMs that result in inactivity of the system with respect to the objective reaction if removed {\cite{clark2012minimal}}. Therefore, they can be used to identify target genes and repress undesirable metabolic functions, whilst assessing the effect on the structure of the entire metabolic network. Elementary flux patterns (EFPs) define all potential elementary routes for steady-state fluxes as sets of indices, and can be mapped to EFMs to include factors such as pathway interdependencies, thus taking the entire network into account {\cite{kaleta2009can}}. Frameworks which combine various computational approaches for synthetic pathway prediction, such as GEM-Path {\cite{campodonico2014generation}} and BNICE {\cite{hatzimanikatis2005exploring}} are increasing in number as they provide the opportunity to calculate all possible paths and score them by efficiency {\cite{medema2012computational}}.

Unbiased methods may also incorporate Monte Carlo sampling, message passing algorithms {\cite{braunstein2008estimating}} or symbolic flux analysis {\cite{schryer2011symbolic}}. The Markov chain Monte Carlo (MCMC) method can be used to uniformly sample metabolic networks from a genotype space, producing a sequence of viable genotypes (or reaction subsets) by performing a reaction swap between each genotype and its successor; if a swap results in a non-viable genotype, this sequence will remain at the previous genotype for that step and the process is repeated until a metabolic network with the correct number of reactions is reached {\cite{samal2010genotype}}. A Monte Carlo based technique has also been used for uniform sampling of feasible steady states in an ellipsoid representing the solution space for a genome-scale metabolic model of \textit{Escherichia coli} \cite{de2013montecarlo}. A revision of this method was proposed with rounding procedures to improve performance by eliminating ill-conditioning when sampling convex polytopes of steady states {\cite{de2015uniform}}.
Of all omic data types, metabolomic data are said to give the closest indication of observed phenotypes {\cite{aurich2016metabotools}}. Therefore, extracellular metabolomic measurements can help to predict intracellular flux states by integrating these data into a constraint-based framework using a sampling-based network approach \cite{mo2009connecting,aurich2015prediction}. The MetaboTools toolbox provides a workflow for integrating metabolomic data into multi-omic models and predicting metabolic phenotypes through analysing how metabolite uptake and secretion differ between conditions {\cite{aurich2016metabotools}}.

\subsection{Biased methods}

Biased methods rely on the definition of an objective function to solve the metabolic network and find its flux rates. For instance, standard flux balance analysis belongs to this class of methods.

\subsubsection{Flux balance analysis and its variants}
\label{subsec:fba}

Among the biased methods, the most well-known technique is flux balance analysis (FBA), which uses the assignment of stoichiometric coefficients in a matrix to represent the metabolites involved in any given reaction in a metabolic network {\cite{orth2010flux}}. Through these coefficients, constraints can be imposed on the system to identify all potential flux distributions associated with a corresponding set of feasible phenotypic states. The aim of FBA is to locate a value (or set of values) in the solution space that best satisfies a given objective function. FBA uses linear programming to solve this objective function, indicating the extent to which each reaction in the network contributes to a phenotypic state. 

If two objectives (flux rates or linear combinations thereof) were to be maximised, a multi-level linear problem would be formulated as follows:
\begin{equation}
\begin{aligned} 
& {\mbox{max}} 
& & g^\intercal v \\
&  {\mbox{such that}}
& & \mbox{max } f^\intercal v \\
& & & {\mbox{such that}} \quad Sv = 0 \\
& & & v^{\mbox{min}} \leq v \leq v^{\mbox{max}} 
\end{aligned}
\label{eq:FBA}
\end{equation}
where $f$ and $g$ are n-dimensional arrays of weights associated with the first and second objectives respectively, and indicate the contribution of the reaction fluxes $v$ to each objective. $v^{\text{min}}$ and $v^{\text{max}}$ are vectors representing the lower and upper limits for the flux rates in $v$.  A constraint in FBA postulates that the total amount of any metabolite being produced must be equal to the total amount of that metabolite consumed {\cite{zielinski2016metabolic}}. The most common objective function computed by FBA is the synthesis of biomass, which is commonly used to indicate cellular growth rate and predict product yields {\cite{feist2010biomass}}. Fluxes can either be calculated under the steady state assumption or in a dynamic state, where changes in specific concentrations and kinetics parameters have been recorded for each metabolite over time (e.g.\ for DFBA) {\cite{mahadevan2002dynamic}}. Experimental validation of model predictions for DFBA are often obtained from $^{13}$C metabolic flux analysis {\cite{wiechert200113}}, which utilises isotopic-labelling of metabolic substrates to quantify intracellular fluxes. Additionally, methods such as dynamic multi-species metabolic modelling (DMMM) have been used to examine inter-species competition for metabolites in a microbial community {\cite{zhuang2011genome}}.

Numerous modifications of FBA propose the application of various constraints to shrink the solution space for determining the precise flux state of the cell by calculating the optimal set of solutions for a given objective via linear programming. In most instances, constraints are defined by cell and reaction stoichiometry, fluxes through transport and metabolic reactions, upper and lower bounds for each flux, biomass composition and ATP requirements {\cite{reed2012shrinking}}. Upper and lower bounds can be estimated using flux variability analysis {\cite{burgard2001minimal}}, which returns the maximum and minimum fluxes through each reaction whilst maintaining minimal biomass production {\cite{mahadevan2003effects}}.
 
Linear thermodynamic constraints can be applied in thermodynamic metabolic flux analysis (TMFA) (or thermodynamics-based flux balance analysis) and thermodynamic variability analysis to eliminate thermodynamically infeasible reactions or loops from pathways and gather information on feasible metabolite activity and Gibbs free energy changes {\cite{henry2007thermodynamics, ataman2015heading}}. The removal of thermodynamically infeasible loops is necessary to prevent violating the loop law, which states that there is no net flux through balanced biochemical loops in networks at steady state {\cite{price2002extreme}}. Loopless COBRA methods solve a modified mixed-integer problem with the added constraint of no network fluxes containing loops; application of this constraint has been described for FBA, FVA and MCMC sampling {\cite{schellenberger2011elimination}}. FBA with thermodynamic constraints has also been described in energy-balance analysis (EBA) {\cite{beard2002energy}}. Fast flux variability analysis with thermodynamic constraints (tFVA) removes unbounded fluxes from biochemical loops arising from non-zero, steady-state fluxes involving internal reactions {\cite{muller2013fast}}. This is a faster implementation of FVA that does not require the specification of metabolite concentrations or additional experimental data, although these have been included in other variants of the method. 

Parsimonious FBA (pFBA) identifies a subset of genes contributing to maximising the growth rate \textit{in-silico}, therefore enabling maximisation of stoichiometric efficiency {\cite{lewis2010omic}}. Another technique uses conditional dependencies present in the metabolic model as constraints for each flux, whereby each flux is constrained by the activity of the compound that facilitates it. This technique is known as conditional FBA and has proved to be effective for simulating phototrophic growth and diurnal dynamics in cyanobacteria \cite{rugen2015elucidating,reimers2016evaluating}. Resource balance analysis (RBA) uses growth rate limitation caused by distribution of proteins between cellular processes to constrain flux predictions {\cite{goelzer2011bacterial}}. Constrained allocation flux balance analysis (CAFBA) applies a genome-wide constraint on fluxes to observe proteome allocation between ribosomal, transport and biosynthetic proteins {\cite{mori2016constrained}}. For this method, growth laws governing the synthesis of intracellular proteins are used to design parameters for predicting levels of protein expression and energy production. To improve the prediction of internal fluxes, cost reduced sub-optimal FBA (corsoFBA) minimises protein and thermodynamic costs to simulate a sub-optimal state {\cite{schultz2015predicting}}. Linear metabolite dilution flux balance analysis (limed-FBA) forces dilution in metabolites associated with growth in active reactions, by adding a small dilution flux to block metabolic pathways without input fluxes {\cite{dreyfuss2013reconstruction}}.

Although more time-consuming than a linear programming approach, Bayesian flux estimation results in a probability density function, which is more stable and informative than a simple point estimate {\cite{heino2007bayesian}}. The METABOLICA statistical framework utilises a Bayesian approach to performing FBA. Metabolism is modelled in a multi-compartment macroscopic model with a stochastic extension of the stationary state and the Bayesian inference problem is solved by computing posterior probability densities using MCMC sampling {\cite{heino2010metabolica}}. 

Alternatively, standard constraints may be removed to construct new FBA methods to improve flux predictions for non-steady-state or non-wild-type cells. Relaxing the assumption of fixed reactant proportions for biomass production is the basis of flexible FBA (flexFBA), which can be coupled with relaxing the fixed ratio of byproduct to reactant (known as time-linked FBA) to observe transitions between steady states {\cite{birch2014incorporation}}. Combining these methods also enables the comparison of metabolite production between knockout mutants. 

To further increase the informativity of constraint-based models, the integration of information from regulatory pathways and external multi-omic data is described in the following section.

\subsubsection{Regulatory methods to generate context-specific metabolic models}
\label{subsec:regulatory}

Regulatory methods can be used to set constraints for FBA which incorporate regulatory networks, as well as introducing external omic data, which provides the opportunity to simulate metabolism under specific genetic or environmental conditions.
Steady-state regulatory flux balance analysis (SR-FBA) is used to quantify the extent to which metabolic and transcriptional regulatory constraints affect the state of flux activity for various metabolic genes {\cite{shlomi2007genome}}. SR-FBA allows for improved characterisation of steady-state metabolic behaviour compared to regulatory FBA (rFBA), which chooses a single steady state per time interval from all possible solutions to find the flux distribution consistent with the regulatory state of each interval. Integrated FBA (iFBA) incorporates metabolic, regulatory and signalling pathways in the FBA model to enable thorough characterisation of dynamic-state metabolic behaviour {\cite{covert2008integrating}}. Integrated dynamic flux balance analysis (idFBA) additionally couples fast and slow reactions to give quantitative time-variant flux predictions {\cite{lee2008dynamic}}. 

However, many of these approaches are limited by the Boolean logic formalism, which restricts the definition of gene activity to an on/off state. This disadvantage can be overcome by using conditional probabilities to represent gene states and gene-transcription factor interactions when combining high throughput data with regulatory networks, as demonstrated by PROM {\cite{chandrasekaran2010probabilistic}}. This approach allows for a greater number of interactions between metabolic models and their respective transcriptional regulatory networks to be recorded as they are quantified automatically {\cite{chandrasekaran2013metabolic}}. Other methods take a differential (rather than absolute) approach to gene expression analysis where gene expression levels classified as belonging to one of three states: over-expressed, unchanged or under-expressed {\cite{rezola2014advances}}. FlexFlux jointly analyses multi-state regulatory networks and metabolic pathways, both of which contribute to calculation of flux {\cite{marmiesse2015flexflux}}. Upon constructing an initial regulatory network, qualitative states evolving towards an `attractor' set are converted into user-defined continuous intervals, thus allowing reactions to be constrained by multiple flux values, rather than one single value. The multi-formalism interaction network simulator (MUFINS) provides a platform for combining multiple kinetic models with signalling and regulatory networks, omic data integration algorithms and steady state FBA with linear inhibitor and activator constraints {\cite{wu2016mufins}}. In this method, a quasi-steady state Petri net (QSSPN) is used to illustrate interactions between different networks {\cite{fisher2013qsspn}}.

Transcriptional-controlled FBA (tFBA) uses constraints between pairs of conditions based on gene expression data for the optimisation of FBA, considering both fold change and absolute change in expression to minimise noise {\cite{van2011predicting}}. Recently, a new method called transcriptional regulated flux balance analysis (TRFBA) has been introduced for incorporating expression data as well as transcriptional regulatory networks to simulate growth under various environmental and genetic perturbations {\cite{motamedian2017trfba}}. TRFBA applies two unique linear constraints. Firstly, reaction rates are limited using a constant that sets expression levels equal to the upper bounds of reactions; secondly, the expression level of each gene is correlated with the expression of the regulating genes. One important advantage of TRFBA is the ability to improve predictions of growth without requiring detailed information about transcriptional regulators and their target genes.

For the integration of transcriptomic profiles with metabolic networks, gene expression measurements can be obtained from microarray and/or RNA sequencing data stored in public repositories such as the Gene Expression Omnibus (GEO) {\cite{barrett2013ncbi}} or Array Express {\cite{kolesnikov2014arrayexpress}}, in order to examine gene activity across various conditions. In addition to these primary archives, there are also databases with additional data processing and annotation {\cite{rung2013reuse} (such as information regarding gene regulation or differential expression under various conditions) e.g.\ Gene Expression Atlas {\cite{petryszak2015expression}}, or the web servers Gene Chaser {\cite{chen2008genechaser} and Profile Chaser {\cite{engreitz2011profilechaser}}, which query GEO. There are also many specialised databases providing functional genomic data relating to a particular disease, species or tissue-type, such as Oncomine (for cancer-specific microarrays) {\cite{rhodes2004oncomine}, the MGI Mouse Gene Expression Database (GXD) {\cite{finger2017mouse}} or the Pancreatic Expression Database (PED) {\cite{ullah2014pancreatic}}. The generation of context-specific metabolic models may be divided into two main approaches: switch-based and valve-based methods. 

\textbf{Switch-based methods for omic integration.} Switch-based algorithms remove inactive or lowly expressed genes by setting the corresponding reaction boundaries to zero before FBA is performed {\cite{salehzadeh2014computational}}. For instance, the algorithm for gene inactivity moderated by metabolism and expression (GIMME) finds a flux distribution which optimises a given objective and avoids the use of so-called `inactive' reactions below a certain transcription threshold {\cite{vivek2016advances,kim2014methods}}. The main advantage of GIMME is that it can re-enable flux associated with false negative values in inactive reactions and record consistency between gene expression data and the predicted flux distribution for a given objective {\cite{becker2008context}}. Gene inactivation moderated by metabolism, metabolomics and expression (GIM3E), is an extended version of GIMME which also incorporates metabolomic data in the form of turnover metabolites added as products to each reaction, along with a corresponding sink reaction {\cite{schmidt2013gim3e}}. This allows for the computation of turnover flux ranges for metabolites.

Tissue-specific gene and protein expression values can be integrated into genome-scale metabolic models accounting for different metabolic objectives at the cellular level {\cite{shlomi2008network}}, in order to extract information regarding the uptake and secretion of metabolites by specific tissue and cell-types. For this, tissue-specific variations in enzyme expression levels are used to inform the likelihood of enzymes supporting flux in their associated reactions by categorising gene-to-reaction mapping for each reaction in the model corresponding to the level of gene expression (i.e. high, moderate or low expression). Subsequently, fluxes corresponding to high gene expression are maximised and those corresponding to low gene expression are minimised when solving a mixed-integer linear program {\cite{blazier2012integration}}. This process has been developed into the Integrative Metabolic Analysis Tool (iMAT) {\cite{zur2010imat}}, which displays the most likely predicted metabolic fluxes corresponding to reactions in metabolic models. This tool enables the definition of a biological objective to be dependent on the requirements of each cell rather than the entire organism. An extension of iMAT known as the exploration of alternative metabolic optima (EXAMO) enables the design of condition-specific metabolic models for human tissues {\cite{rossell2013inferring}}.

Similarly, the integrative network inference for tissues (INIT) algorithm uses tissue-specific information collected from the Human Protein Atlas to help incorporate transcriptomic and proteomic data into a genome-scale model and produce cell-type specific metabolic networks {\cite{agren2012reconstruction}}. This data forms the input for a mixed-integer linear problem, which modifies the steady-state condition by setting a small positive net accumulation rate for internal metabolites {\cite{machado2014systematic}}. Net productions of these metabolites are assigned positive weights, corresponding to arbitrary scores for the level of protein expression. An updated version of INIT known as the task-driven integrative network inference for tissues (tINIT) was devised to identify structural analogs to metabolites (so-called `antimetabolites') and a core set of metabolic tasks to be included in the model {\cite{agren2014identification}}. tINIT prevents simultaneous flux in reversible reactions and allows the user to decide whether net production of all metabolites should be considered.
The algorithm for metabolic adjustment by differential expression (MADE) compares the fold changes of gene expression values between conditions to intuitively predict the most consistent and statistically-significant metabolic adjustments {\cite{jensen2011functional}}. The fold changes are expressed as a series of binary expression states, for which differences between successive states most closely mirror corresponding differences in the mean expression levels. 

\textbf{Valve-based  methods for omic integration.} Unlike switch-based methods, valve-based algorithms reduce the activity of lowly expressed genes by adjusting the upper and lower bounds for their corresponding reactions. This is usually proportional to the normalised expression of the genes associated with those reactions before performing FBA {\cite{salehzadeh2014computational}}. Such methods include E-flux {\cite{colijn2009interpreting}}, E-flux2 {\cite{kim2016flux2}}, METRADE {\cite{angione2015predictive}}, FALCON {\cite{barker2015robust}} and PROM {\cite{chandrasekaran2010probabilistic}}. For valve-based methods, gene expression data is not discretised as in switch-based methods. Data from methods that treat gene expression as relative as opposed to absolute are more indicative of protein concentrations, as levels of transcription are more comparable across genes {\cite{lee2012improving,machado2014systematic}}. In E-flux, flux boundaries are tightly constrained when gene expression is low but relaxed when gene expression is high {\cite{colijn2009interpreting}}; transcript levels can be used to set an upper bound for the maximum production of enzymes and therefore constrain all reaction rates {\cite{kim2014methods}}. 
To improve this formulation, E-flux2 adds minimisation of the Euclidean norm of the measured flux vector, thus generating a unique solution {\cite{kim2016flux2}}. 

Personalised reconstruction of metabolic models (PRIME) creates cell-specific models incorporating both transcriptomic and phenotypic data, and only modifies the bounds of a small set of reactions within a pre-defined range {\cite{yizhak2014phenotype}}. Expression data-guided flux minimisation (E-Fmin) is similar to GIMME in that it minimises a sum of fluxes where weights are a function of gene expression level; however, biomass production is forced to carry non-zero flux (i.e. metabolic activity is not threshold-dependent) and all  reactions are thermodynamically-feasible due to flux minimisation {\cite{song2014prediction}}. FALCON is a novel algorithm that estimates enzyme abundances using gene-protein-reaction (GPR) rules in the model, thus improving the predictive capability of models integrated with expression data {\cite{barker2015robust}}. Within a multi-omic model, the metabolic and transcriptomics adaptation estimator (METRADE) constructs a Pareto front in order to identify the best trade-off when multiple objectives are simultaneously optimised {\cite{angione2015predictive}}. Transcriptomic data comprising gene expression profiles and codon usage arrays can be mapped to a phenotypic objective space where each profile is associated with a condition {\cite{kashaf2017making}}. In this way, the identification of optimal metabolic phenotypes is facilitated through the concurrent maximisation or minimisation of multiple metabolic markers and comparison of predicted flux rates between objectives.

\textbf{Network pruning.} In addition to switch- and valve- based integration methods, there are also pruning methods such as MBA {\cite{jerby2010computational}}, FASTCORE {\cite{vlassis2014fast}}, mCADRE {\cite{wang2012reconstruction}} and OnePrune {\cite{dreyfuss2013reconstruction}}, which only retain a core set of reactions in the metabolic model. FASTCORMICS is a faster adaptation of FASTCORE which facilitates data integration by pre-processing and produces multiple metabolic models {\cite{pacheco2015integrated}}. Similarly, the cost optimisation reaction dependency assessment (CORDA) algorithm performs a four-step dependency assessment before calculating flux whilst minimising cost production i.e. utilising as many high confidence reactions as possible and minimising the involvement of absent reactions {\cite{schultz2016reconstruction}}. The cost of each reaction in the network is represented by the addition of a pseudo-metabolite as a product. CORDA is quicker to implement than many other pruning methods owing to its use of FBA to calculate flux.

On the other hand, there are many methods do not fit into the aforementioned categories (switch/valve/network pruning) as they utilise more unconventional approaches for omic data integration. Similar to E-flux2, the regularised context-specific model extraction method (RegrEx) is based on principles of regularised least squares optimisation by minimising the squared Euclidean distance between fluxes and experimental data {\cite{estevez2015context}} to calculate fluxes which are independent of user-defined parameters. Instead of assigning expression measurements to individual genes or reactions, the GC-Flux algorithm splits GPR strings for each reaction into functional gene complexes to overcome the assumption of proteins catalysing more than one reaction at a time {\cite{fyson2017gene}}. Another recent development is the use of the Huber penalty convex optimisation function (HPCOF) combined with flux minimisation, to achieve a more accurate prediction of fluxes which are closer to experimentally measured values {\cite{zhang2017prediction}}. This method introduces continuous gene expression values in the form of both constraints and target equations, without the need for definition of a biomass objective function or expression thresholds. A novel method known as omFBA {\cite{guo2016om}} uses a phenotype-match algorithm to formulate the optimal objective function, i.e. the function that yields the most accurate estimations of the observed phenotypes. This objective can be simultaneously correlated with multiple omic data types via regression analysis to generate a omics-guided objective function, consequently resulting in a clearer correlation between genotype and phenotype and improved phenotypic predictions.

\subsection{Genetic perturbation and objective function selection}
\label{subsec:perturbation}

Genetic perturbation is an important tool for establishing gene essentiality and maximising pathway efficiency. Deciding upon the number of gene perturbations to be performed depends on multiple factors. Choosing to perform single or pairwise gene perturbations one-by-one may fail to capture the essentiality and function of that gene as a result of genetic redundancy (i.e. there may be multiple genes encoding the same function). However, concurrently knocking out multiple genes can cause issues related to scaling unless coupled with e.g.\ Shapley value analysis, which assigns a contribution value to each gene knockout in the system {\cite{deutscher2008can}}. Synthetic lethality can be described as the simultaneous inactivation of a set of non-essential genes resulting in the death of a cell or organism {\cite{nijman2011synthetic}}. Knocking out multiple synthetic lethal pairs for genome-scale metabolic models can help in analysing the structural robustness of metabolic networks and identifying interdependencies among genes and reactions {\cite{suthers2009genome}}. 

There are numerous algorithms for the detection and analysis of synthetic lethal pairs. Fast-SL is an algorithm capable of identifying higher order lethal reaction and gene sets by taking GPR (gene-protein-reaction) associations into account and vastly reducing the search space before iterating through the remaining combinations of genes/reactions {\cite{pratapa2015fast}}. Minimal cut sets can also be regarded as synthetic lethals, which can be targeted for drug therapies as they constitute essential gene/reaction sets {\cite{tobalina2016direct}}. The data mining synthetic lethality identification pipeline (DAISY) statistically infers interactions between synthetic lethals using a combination of approaches: genomic survival of the fittest (SoF), shRNA-based functional examination and pairwise gene coexpression {\cite{jerby2014predicting}}. A method for identifying dosage lethality effects (IDLE) in genome-scale models of cancer metabolism exploits synthetic dosage lethality to simulate the pairwise knockout of non-essential enzymes via overexpression of the first gene and underexpression of the second {\cite{megchelenbrink2015synthetic}}.

Flux ratios can be applied as constraints for FBA using the flux balance analysis with flux ratios (FBrAtio) algorithm, which can be directly implemented into the stoichiometric matrix of genome-scale metabolic models {\cite{mcanulty2012genome,yen2013deriving}}. In FBrAtio, multiple enzymes compete for metabolic branch points in the network (known as critical nodes) {\cite{mcanulty2012genome}}, which specify how a substrate in the metabolite pool is distributed between competing reactions; this depends on factors  relating to thermodynamics such as enzyme availability and downstream accumulation of reactive intermediates. The optimisation of flux ratios for a particular phenotype can be achieved through partial knockdown, overexpression or total knockout of enzyme-coding genes {\cite{mcanulty2012genome}}. As opposed to complete knockouts, performing gene over-expression or partial knockdown may prove to be useful for targeted reduction of expression levels. 

The minimisation of metabolic adjustment (MOMA) method relaxes the assumption of optimal growth flux for gene deletions by solving a quadratic problem to optimise distance minimisation in flux space {\cite{segre2002analysis}}. This is because the minimal response to the perturbation is considered to be a more accurate estimate of the true flux state in the mutant {\cite{raval2013introduction}}. Initially, the flux distribution for the mutant remains as close as possible to optimal flux for the wild-type and deviates to form a sub-optimal flux distribution between that of the wild-type and mutant. In this way, MOMA is able to predict phenotypic outcomes following knockouts more precisely than FBA. 

Using a mechanistic model of reaction rates, integrative omics metabolic analysis (IOMA) also solves a quadratic problem to deliver kinetically-derived estimations of flux following genetic perturbation {\cite{yizhak2010integrating}}. This is possible through integrating quantitative proteomic and metabolomic data into the model, which improves performance when compared to MOMA. Similarly, regulatory on/off minimisation (ROOM) minimises the number of significant flux changes following knockouts with respect to the wild type {\cite{shlomi2005regulatory}}. This is achieved through redirecting flux through alternative pathways following knockout. 

Many gene perturbation experiments simulate knockouts under the assumption that there is no downstream effect on gene regulation {\cite{lee2012improving}. In the RELATCH method, the principle of relative optimality is applied to predict how cells adapt to perturbations, by minimising relative flux patterns and latent pathway activation with respect to a reference flux distribution {\cite{kim2012relatch}}. As strains adapt to their perturbed state, they undergo regulatory and metabolic changes, represented by two parameters - one penalising latent pathway activation and another limiting enzyme contribution increases in active pathways. In varying environmental conditions, Bayesian factor modelling can be used to elucidate pathway cross-correlations and identify degrees of pathway activation \cite{angione2015hybrid}. Unconventionally, the REMEP method considers the impact of perturbations on metabolite as well as flux patterns {\cite{oyetunde2017metabolite}}. This leads to improved flux predictions for knockout mutants as the structure of cellular regulation is represented more accurately.

\section{Multi-objective optimisation of metabolic models}
\label{sec:optimisation}

It can be difficult to define the single most important objective in a biological system as there are usually multiple conflicting cellular objectives in addition to the maximisation of biomass, which is often used as a proxy for growth. Methods such as Bayesian objective function discrimination can be used for selection of the most suitable objective function by using a probabilistic approach to compare multiple objectives {\cite{knorr2007bayesian}}. Alternatively, there are methods utilising lexicographic ordering {\cite{sendin2010multi}} or calculation of a weighted sum to scalarise multiple objectives {\cite{xu2011iterative}}; however, it can be difficult to select weights that elicit a uniform distribution of Pareto solutions and find solutions in non-convex regions {\cite{angione2015multi,de2014global,angione2013pareto}}. Thus, multi-objective optimisation arguably presents the most realistic representation of metabolic flux in biological systems by considering the contribution of a wide range of competing objectives. The rest of the section describes the main methods used to implement multi-objective optimisation in metabolic models. 

Multi-objective optimisation can be used to resolve trade-offs between conflicting metabolic objectives through simulating a series of optimal, non-dominated vectors \textit{f(x)} in the multi-dimensional objective space. For such vectors, an improved solution does not exist for any given objective without sacrificing the performance of another {\cite{angione2015predictive}}. This is known as a Pareto front and enables the simultaneous consideration of multiple conditions and constraints affecting each cellular objective {\cite{xu2009multi}}. For example, optimisation of the objective function \textit{r} through maximisation or minimisation of the vector function \textit{f(x)} may be carried out respectively as follows:
\begin{equation}
\begin{aligned} 
f_i (x) > f_i (x^\ast), \;\forall \; i = 1,...,r \\
f_i (x) < f_i (x^\ast), \;\forall \; i = 1,...,r
\end{aligned}
\label{eq:Pareto_optimal}
\end{equation}
where $x\textsuperscript{*}$ constitutes all non-dominated vectors present in the search space, for which there is no point \textit{x} such that either of the above statements are satisfied (depending on whether a maximisation or minimisation is carried out). 
On the other hand, the non-linearity of metabolic networks means that there is often concavity or discontinuity present in Pareto fronts {\cite{sendin2006model}}. These issues may be resolved through the use of multi-objective evolutionary algorithms (MOEAs), through which it is possible to obtain the entire set of Pareto-optimal solutions by running the algorithm only once \cite{angione2015analysis}. Genetic design by multi-objective optimisation (GDMO) \cite{costanza2012robust} employs MOEAs to find genetic manipulations (in the form of Pareto-optimal solutions) that simultaneously optimise multiple metabolic objectives. 

The most widely used MOEA is NSGA-II {\cite{deb2002fast}}, which conducts cross-comparisons between points in the objective space to establish whether a higher value exists for all objectives. This process is known as non-dominated sorting, as it involves categorising values in the distribution as either dominated or non-dominated. The non-dominated values are ordered into a front and the normalised distances between these points and their nearest neighbours are computed for each front. These are known as crowding distances and are necessary to preserve diversity (i.e. obtain a good spread of solutions), but may cause instability if two or more points share the same fitness values {\cite{fortin2013revisiting}}. A sphere-excluding evolutionary algorithm (SEEA) has been proposed for maximising diversity in NSGA-II and preventing convergence at local optima {\cite{zheng2012enhancing}}. Through comparing the non-dominated fronts, it is possible to establish the prioritisation of objectives for the production of specific metabolites. 

In the interests of metabolic engineering, it is often useful to ascertain which knockouts would optimise the production of a specific metabolite. Commonly used for strain improvement in metabolic engineering, OptKnock is a computational method that identifies and subsequently removes metabolic reactions capable of uncoupling biomass maximisation from the production of a specific metabolite using a nested optimisation framework {\cite{burgard2003optknock}}. Cellular transport rates and secretion pathways can also be used to further constrain this multi-objective model {\cite{pharkya2003exploring}}. RobustKnock is an extension of OptKnock that considers the role of competing pathways in diverting metabolic flux away from production of the desired metabolite, thus leading to sub-optimal flux distributions {\cite{tepper2010predicting}}. Therefore, the removal of these competing pathways and improved knockout strategies result in more robust flux predictions. OptForce can be used to specify perturbations leading to targeted overproduction of a metabolite {\cite{ranganathan2010optforce}}, with a variant known as k-OptForce which includes kinetic parameters if available (in the form of kinetic rate laws and metabolite concentrations) {\cite{chowdhury2014k}}. ReacKnock proposes an improved solution to previous methods defining strategies for deleting reactions, in that the Karush-Kuhn-Tucker (KKT) method is used to reformulate the problem for single-level optimisation {\cite{xu2013reacknock}}. As well as reaction deletions, OptStrain provides strategies for reaction addition through identifying non-native reactions in universal databases that are likely to improve product yields {\cite{pharkya2004optstrain}}. Furthermore, SimOptStrain considers gene deletions, addition of non-native reactions and gene-protein-reaction rules for optimising a given pathway {\cite{kim2011large}. Other strategies for gene deletion include genetic design through local search (GDLS) {\cite{lun2009large}}, the cipher of evolutionary design (CiED) {\cite{fowler2009increased}}, simulated annealing (SA) algorithms and  set-based evolutionary algorithms (SEAs) {\cite{rocha2008natural}}.

Linear Physical Programming-Based Flux Balance Analysis (LPPFBA) aims to prioritise objectives and constraints for a given set of Pareto-optimal solutions, thus aiding the identification of conflicting objectives and regions of the solution space that contain feasible optimal fluxes {\cite{nagrath2010soft}}. Comprehensive Polyhedra Enumeration Flux Balance Analysis (CoPE-FBA) indicates the topology of the sub-networks corresponding to optimal flux vectors in polyhedra with an emphasis on vertices of the solution space {\cite{kelk2012optimal}}. This process can be refined by dividing reversible reactions (termed linealities) into separate forward and backward reactions, thus simplifying the optimal solution space for optimisation of more objectives and yielding all non-decomposable flux routes {\cite{maarleveld2015interplay}}.

Metabolic interactions between species in a microbial community have been simulated with respect to multiple objectives using algorithms such as OptCom, which provides a framework for the comparison of fitness trade-offs for individual species against that of the community {\cite{zomorrodi2012optcom}}. Community flux balance analysis (cFBA) utilises non-linear multi-objective optimisation to build a more complete picture of metabolic fluxes by predicting biomass abundance as well as metabolite exchanges with the addition of community-specific constraints {\cite{khandelwal2013community}. However, only flux distributions resulting in optimisation of the community growth rate are identified and the quality of flux predictions obtained using this method is heavily reliant on the quality of model reconstructions {\cite{gottstein2016constraint}}. The community and systems-level interactive optimisation (CASINO) toolbox has been developed to conduct multi-level optimisation and phenotypic prediction for analysis of microbial interactions within a metabolic model of the human gut {\cite{shoaie2015quantifying}}. Here, biomass production for individual species as well as the microbial community as a whole is calculated, starting from a community matrix defining the topology of all reactions. The microbial community modeller (MCM) is another such tool where parametric fitting, sensitivity analysis and statistical evaluation are incorporated into models to assess the metabolic potential of each species in a community at the cellular level {\cite{louca2015calibration}}. The computation of microbial ecosystems in time and space (COMETS) takes spatio-temporal dynamics into account by simulating time-dependent fluxes on a lattice containing information about the spatial distribution of microbial species and nutrients within a community; in addition to interactions, growth and uptake of substrates by different species can be simulated to examine how intracellular resource allocation is locally optimised by each species {\cite{harcombe2014metabolic}}. 

For multi-objective optimisation in dynamic flux states, changing intra- and extracellular concentrations of metabolites (and their corresponding gene expression values) can be observed over time using MetDFBA {\cite{willemsen2015metdfba}}, which reduces the number of parameters for DFBA, thereby improving flux estimations. In this manner, different objective functions representing various phenotypes can be compared, and objectives that constrain fluxes corresponding to specific metabolites can be noted so that the feasible solution space can be further constrained. d-OptCom is a dynamic, multi-level extension of OptCom through which biomass accumulation and exchange of metabolites in microbial communities can be examined using dynamic mass balance equations and substrate uptake kinetics {\cite{zomorrodi2014d}}. The AMIGO2 Toolbox reformulates optimal control problems as dynamic multi-objective optimisation problems, which can be solved using non-linear programming {\cite{balsa2016amigo2}}. To circumvent the difficulties of acquiring time-course data to generate detailed kinetic models, ensemble modelling methods utilise steady state phenotypic data (such as flux changes caused by genetic perturbations) to specify a set of models which represent the dynamics of a metabolic system {\cite{tran2008ensemble}}. Ensemble Modeling for Robustness Analysis (EMRA) considers how to maintain robustness in non-native pathways by building an ensemble of models linked to the same steady-state flux distribution. Subsequently, a continuation approach is used to alter kinetic parameters until a bifurcation point is found, where the steady state disappears {\cite{lee2014ensemble}}.

Noninferior set estimation (NISE) has been used to divide the solution space and compute weights for multiple objectives in a multi-objective method for FBA (MOFBA). MOFBA can (i) generate Pareto curves demonstrating competing metabolic objectives and (ii) compute individual flux distributions for each Pareto optimal solution {\cite{oh2009multiobjective}}. Large scale differential algebraic equation (DAE) solvers can be used to construct Pareto optimal curves for various perturbations such as heat shock, and observe where the nominal operating point for the wild type lies with respect to each curve {\cite{el2005optimal}}. There are variations of DAE systems that allow for sensitivity analysis to examine the rates of changes caused by small perturbations. Usually, sensitivity analysis examines the effect of changes in the reaction {\cite{stracquadanio2010analysis}}, pathway {\cite{costanza2012robust}} and species {\cite{zhang2010comparison}} spaces of metabolic models {\cite{petzold2006sensitivity}}. The recently developed method for thermodynamics-based metabolite sensitivity analysis (TMSA) ranks metabolites on their ability to limit solutions to thermodynamically-consistent reactions and provides information about thermodynamic uncertainty in metabolic networks {\cite{kiparissides2016thermodynamics}}. ORACLE (optimisation and risk analysis of complex living entities) evaluates quantitative uncertainty in kinetic models by sampling metabolite concentrations, and computing elasticity for enzyme states which represent the displacement of enzymes from thermodynamic equilibrium {\cite{miskovic2010production}}. iSCHRUNK extends the ORACLE approach by including machine-learning classification algorithms to specify enzyme saturation levels and derive a more feasible population of kinetic models than ORACLE through achieving better characterisation of the kinetic parameters in the solution space {\cite{andreozzi2016ischrunk}}.

In the following section, we present a hands-on tutorial for genetic design by multi-objective optimisation. Although several pipelines are available in Matlab and Python, this is the first tutorial available in R for multi-objective optimisation of FBA models.

\subsection{Genetic design by multi-objective optimisation: an R tutorial}
\label{gene-design-by-multi-objective-optimization-an-r-tutorial}

This tutorial on genetic design by multi-objective optimisation (GDMO) \cite{costanza2012robust} shows how a multi-objective genetic optimisation algorithm (in this case, a modification of NSGA-II \cite{deb2002fast}) can be used to optimise the trade-off between multiple metabolic objectives. This is useful both as a tool for metabolic engineering (for example, when finding the favourable set of nutrients to optimise a microbial strain), and for quantifying the relationship between the objectives. The flowchart in Figure \ref{fig:flowchart} provides a brief overview of the main stages of GDMO performed in R. Although we provide an explanation of the code, we suggest the following books for familiarisation with the fundamentals of programming in R and evolutionary optimisation methods {\cite{teetor2011r,matloff2011art,deb2001multi}}.

R, Python and Matlab all have good support for metabolic modelling. For a researcher approaching the field, the primary deciding factor is personal programming language preference. Performance is largely unrelated to language because the most time consuming step is optimising the linear optimisation problem, which is performed by an external linear programming toolkit. We have had particularly good results with the commercial solver Gurobi and the open source solver GLPK.

\begin{figure}[htbp]
\centering
\includegraphics[width=0.8\textwidth]{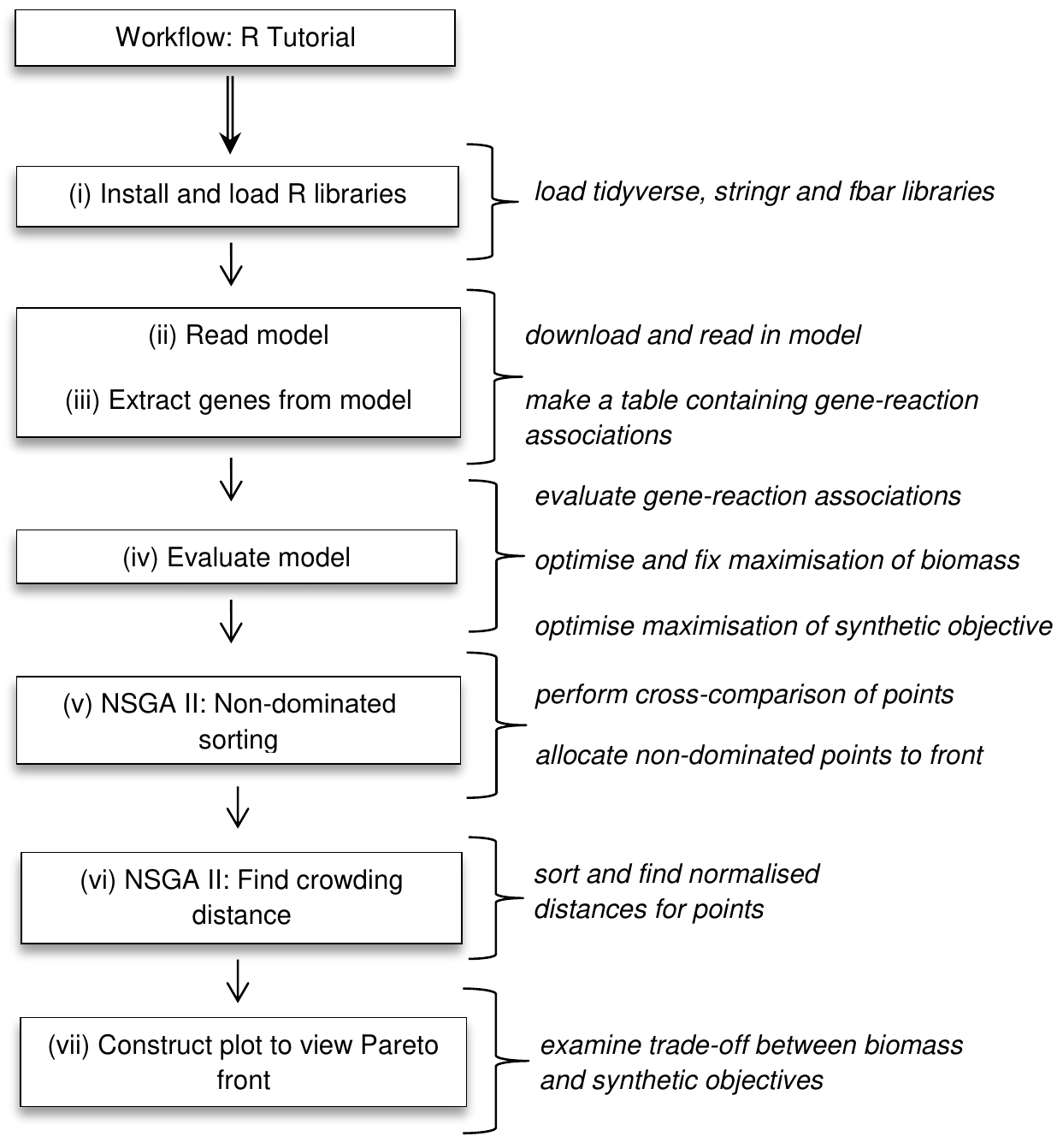}
\caption{A flowchart outlining the main stages of the R tutorial: (i) loading the requisite R libraries for analysis; (ii) reading the metabolic model into R; (iii) compiling a table that associates a list of genes extracted from the model with the reactions they are involved in; (iv) evaluating these associations by checking for the presence of genes in each iteration and performing FBA to obtain estimates of the biomass and synthetic objectives; (v) implementing a custom version of the NSGA-II algorithm, which conducts comparisons between points in the flux distribution to establish whether a higher value exists for all objectives, thus categorising these as dominated and non-dominated, the latter of which are ordered into a front; (vi) computing crowding distance i.e. the normalised distances between non-dominated points and their nearest neighbours in each front and dimension; (vii) viewing the Pareto front, through which it is possible to establish the prioritisation of objectives for the production of specific metabolites. The full code for this tutorial is provided in the Supplementary Material, and is downloadable from \url{http://modellingmetabolism.net}.}
\label{fig:flowchart}
\end{figure}

Loading and preparing a metabolic model are the first steps of any metabolic modelling procedure (see Supplementary Material). Assuming the initial steps have been carried out, we now describe the evaluation function that we use. In the context of multi-objective modelling, the evaluation function returns a number of values, each of which is used as an objective to be optimised. The evaluation function that we use here has four main stages:
\begin{enumerate}
\itemsep1pt\parskip0pt\parsep0pt
\item
  The gene-reaction associations (\texttt{geneAssociation}) are
  evaluated in the context of which genes are present in this iteration
  (\texttt{genome}), to give an \texttt{activation} value, which is an estimate of reaction rate.
\item
  The \texttt{activation} value is used to alter the upper and lower bounds on reaction rate (\texttt{uppbnd} and \texttt{lowbnd}), to push reaction rates towards the rate estimates.
\item
  We conduct a round of FBA, optimising for maximum biomass.
\item
  We fix the biomass production value to its maximum, by altering the corresponding \texttt{uppbnd} and \texttt{lowbnd} to be near the \texttt{flux} (+/-1\%).
\item
  With the biomass value fixed, we alter the objective coefficient (\texttt{obj\_coef}) to target optimisation of the synthetic objective (acetate in the example).
\end{enumerate}

The technique of fixing the biomass followed by maximising the synthetic
objective is important because there could still be slack in the model
after the first optimisation stage, which would allow for multiple possible synthetic objective values for a given biomass value and genome. This slack must be removed by a second optimisation round since we wish to have a correct and consistent estimate of the synthetic objective.

\begin{Shaded}
\begin{Highlighting}[]
\NormalTok{evaluation_function <-}\StringTok{ }\NormalTok{function(genome)\{}
  
  \NormalTok{res <-}\StringTok{ }\NormalTok{model %>%}
\StringTok{    }\KeywordTok{mutate}\NormalTok{(}\DataTypeTok{activation =} \NormalTok{fbar::}\KeywordTok{gene_eval}\NormalTok{(}\DataTypeTok{expressions =} \NormalTok{geneAssociation, }
                                        \DataTypeTok{genes =} \KeywordTok{names}\NormalTok{(genome), }
                                        \DataTypeTok{presences =} \NormalTok{genome}
                                        \NormalTok{),}
           \DataTypeTok{activation =} \KeywordTok{coalesce}\NormalTok{(activation, }\DecValTok{1}\NormalTok{),}
           \DataTypeTok{uppbnd =} \KeywordTok{pmin}\NormalTok{(uppbnd, }\DecValTok{1000}\NormalTok{*activation}\FloatTok{+0.1}\NormalTok{),}
           \DataTypeTok{lowbnd =} \KeywordTok{pmax}\NormalTok{(lowbnd, -}\DecValTok{1000}\NormalTok{*activation}\FloatTok{-0.1}\NormalTok{)) %>%}
\StringTok{    }\NormalTok{fbar::}\KeywordTok{find_fluxes_df}\NormalTok{(}\DataTypeTok{do_minimization =} \OtherTok{FALSE}\NormalTok{) %>%}
\StringTok{    }\KeywordTok{mutate}\NormalTok{(}\DataTypeTok{lowbnd =} \KeywordTok{ifelse}\NormalTok{(abbreviation==}\StringTok{'Biomass_Ecoli_core_w/GAM'}\NormalTok{, }
                           \NormalTok{flux*}\FloatTok{0.99}\NormalTok{, }
                           \NormalTok{lowbnd),}
           \DataTypeTok{uppbnd =} \KeywordTok{ifelse}\NormalTok{(abbreviation==}\StringTok{'Biomass_Ecoli_core_w/GAM'}\NormalTok{, }
                           \NormalTok{flux*}\FloatTok{1.01}\NormalTok{, }
                           \NormalTok{uppbnd),}
           \DataTypeTok{obj_coef =} \DecValTok{1}\NormalTok{*(abbreviation==}\StringTok{'EX_ac(e)'}\NormalTok{)) %>%}
\StringTok{    }\NormalTok{fbar::}\KeywordTok{find_fluxes_df}\NormalTok{(}\DataTypeTok{do_minimization =} \OtherTok{FALSE}\NormalTok{)}
  
  \KeywordTok{return}\NormalTok{(}\KeywordTok{list}\NormalTok{(}\DataTypeTok{bm =} \KeywordTok{filter}\NormalTok{(res, abbreviation==}\StringTok{'Biomass_Ecoli_core_w/GAM'}\NormalTok{)$flux, }
              \DataTypeTok{synth =} \KeywordTok{filter}\NormalTok{(res, abbreviation==}\StringTok{'EX_ac(e)'}\NormalTok{)$flux))}
\NormalTok{\}}
\end{Highlighting}
\end{Shaded}

Before proceeding with the genetic optimisation portion of the procedure, we need to describe two helper functions to our slightly modified form of NSGA-II, termed \texttt{non\_dom\_sort} and \texttt{crowding\_distance}. Note that the following descriptions refer to the full code provided in the complete tutorial, available as Supplementary Material.

Non-domination sorting is the first stage of the selection procedure in
NSGA-II. It sorts the points by multiple objectives to Pareto, or non-dominated fronts. These fronts are designed such that for every point in a front, there is no point in the same front or another front with a higher number such that the second point is better than the first in every objective (see Equation \ref{eq:Pareto_optimal}). This is calculated as follows:
\begin{enumerate}
\itemsep1pt\parskip0pt\parsep0pt
\item
  We compare every point against
  every other point.
\item
  For each point (\textit{x}), we see if there exists any second
  point (\textit{y}) that has a higher value in all
  objectives. Where such a second point exists, we term the original
  point `dominated'. 
\item
  We find the set of points that have no dominating point, and term
  this the first non-dominated front. When two points are identical, they are both assigned to the same front.
\item
  We repeat this procedure, but ignore points in the first
  non-dominated front to find the second non-dominated front, and so on.
\end{enumerate}

The second part of the NSGA-II evaluation procedure is finding the
crowding distance. This is used to break ties between points in the same
non-dominated front. For each front and for each dimension, this
function sorts the points into order along the dimension, and finds the
normalised distance between the proceeding point and succeeding point.
These values are summed up across each dimension to find the value for
the point.

The following code is the genetic loop of the algorithm. It is explained by code comments, but follows a normal pattern of evaluating, sorting, selecting from and mutating the population. The genetic algorithm used here is a modified version of NSGA-II \cite{deb2002fast}, with a population of $200$ individuals and carrying out $500$ iterations. Inside the loop, the steps are as follows:
\begin{enumerate}
\item Evaluate all genomes: first, we use the evaluation function on each genome to find the resulting biomass and synthetic fluxes.
\item Round the results: this is a useful tool to help the NSGA-II procedure by regarding very similar results as identical, encouraging more variety in the results set.
\item Label the results: labelling is required so that we can identify them after non-dominance sorting.
\item Shuffle: shuffling the results is important because inevitably, some points will be completely identical, and we want to choose one at random in this case, rather than always pick the same result.
\item Find the non-dominated fronts: assign a front number to each point, such that points with a lower front number are strictly superior to those with a higher one.
\item Find the crowding distance: select for points with more variety.
\item Sort by front, breaking ties by crowding distance: there are normally multiple points in each front, so the continuous crowding distance value is required to choose between these.
\item Keep the best half of the population: using the label assigned earlier.
\item Sample parents from population: use random sampling with replacement to find which members of the population are used as a basis for new members.
\item Mutate parents to create offspring: add new members to the population by flipping 2\% of the genes in the parents.
\item Combine the offspring and parent populations: build the new population and repeat.
\end{enumerate}

\begin{Shaded}
\begin{Highlighting}[]
\NormalTok{start_genome <-}\StringTok{ }\KeywordTok{set_names}\NormalTok{(}\KeywordTok{rep_along}\NormalTok{(genes_in_model, }\OtherTok{TRUE}\NormalTok{), genes_in_model)}
\NormalTok{pop <-}\StringTok{ }\KeywordTok{list}\NormalTok{(start_genome)}

\NormalTok{popsize =}\StringTok{ }\DecValTok{200}
\NormalTok{generations =}\StringTok{ }\DecValTok{500}

\NormalTok{for(i in }\DecValTok{1}\NormalTok{:generations)\{}
  \NormalTok{results <-}\StringTok{ }\KeywordTok{map_df}\NormalTok{(pop, evaluation_function) %>%}\StringTok{ }\CommentTok{# Evaluate all the genomes}
\StringTok{    }\KeywordTok{mutate}\NormalTok{(}\DataTypeTok{bm=}\KeywordTok{signif}\NormalTok{(bm), }\DataTypeTok{synth=}\KeywordTok{signif}\NormalTok{(synth)) %>%}\StringTok{ }\CommentTok{# Round results}
\StringTok{    }\KeywordTok{mutate}\NormalTok{(}\DataTypeTok{id =} \DecValTok{1}\NormalTok{:}\KeywordTok{n}\NormalTok{()) %>%}\StringTok{ }\CommentTok{# label the results}
\StringTok{    }\KeywordTok{sample_frac}\NormalTok{() %>%}\StringTok{ }\CommentTok{# Shuffle}
\StringTok{    }\KeywordTok{non_dom_sort}\NormalTok{() %>%}\StringTok{ }\CommentTok{# Find the non-dominated fronts}
\StringTok{    }\KeywordTok{crowding_distance}\NormalTok{() %>%}\StringTok{ }\CommentTok{# Find the crowding distances}
\StringTok{    }\KeywordTok{arrange}\NormalTok{(front, }\KeywordTok{desc}\NormalTok{(crowding)) }\CommentTok{# Sort by front, breaking ties by crowding distance}
  
  \NormalTok{selected <-}\StringTok{ }\NormalTok{results %>%}
\StringTok{    }\KeywordTok{filter}\NormalTok{(}\KeywordTok{row_number}\NormalTok{() <=}\StringTok{ }\NormalTok{popsize/}\DecValTok{2}\NormalTok{) %>%}\StringTok{ }\CommentTok{# Keep the best half of the population}
\StringTok{    }\KeywordTok{getElement}\NormalTok{(}\StringTok{'id'}\NormalTok{)}
  
  \NormalTok{kept_pop <-}\StringTok{ }\NormalTok{pop[selected]}
  
  \NormalTok{altered_pop <-}\StringTok{ }\NormalTok{kept_pop %>%}
\StringTok{    }\KeywordTok{sample}\NormalTok{(popsize-}\KeywordTok{length}\NormalTok{(selected), }\OtherTok{TRUE}\NormalTok{) %>%}\StringTok{ }\CommentTok{# Sample parents from population}
\StringTok{    }\KeywordTok{map}\NormalTok{(function(genome)\{}
      \KeywordTok{xor}\NormalTok{(genome, }\KeywordTok{runif}\NormalTok{(}\KeywordTok{length}\NormalTok{(genome))>}\FloatTok{0.98}\NormalTok{) }\CommentTok{# Mutate parents to create offspring}
      \NormalTok{\})}
  
  \NormalTok{pop <-}\StringTok{ }\KeywordTok{unique}\NormalTok{(}\KeywordTok{c}\NormalTok{(kept_pop, altered_pop)) }\CommentTok{# Combine the offspring and parent populations}
\NormalTok{\}}
\end{Highlighting}
\end{Shaded}


Once we have a results set, we can construct a plot like Figure \ref{fig:pareto_front} to view the non-dominated fronts. 
We can see how the first front describes the trade-off between biomass and the synthetic objective, with the lines showing the dominated area (to the bottom left). 
This shows that even with a small population and number of iterations we can see three high quality tradeoff points between the synthetic objective and biomass production, with biomass values of around $0.75$, $0.63$ and $0.38$ h$^{-1}$, and corresponding synthetic values of $6$, $11$ and $14$ mmol h$^{-1}$ gDW$^{-1}$.

\begin{figure}[htbp]
\centering
\includegraphics[width=0.6\textwidth]{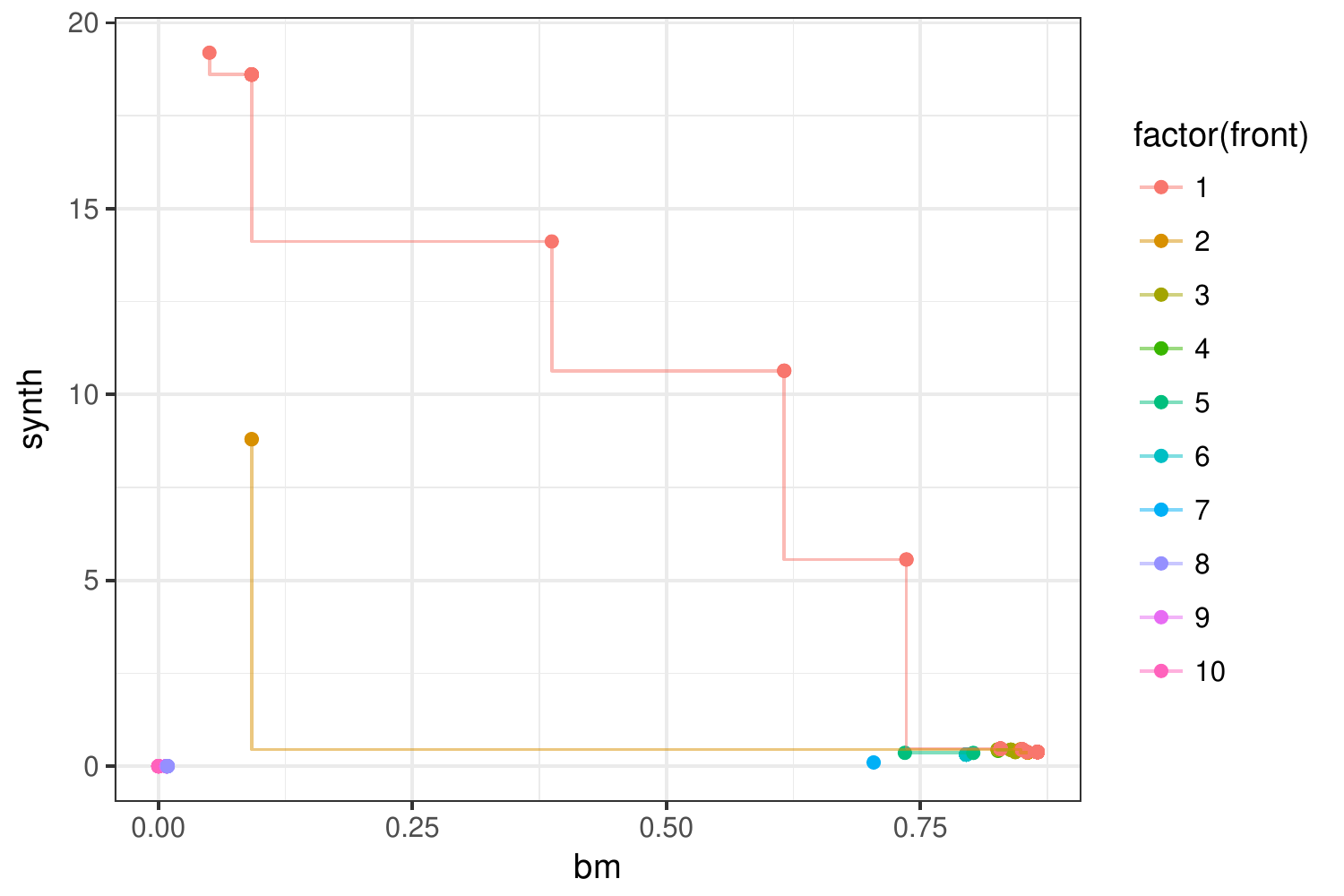}
\caption{Plot of a sample Pareto front showing the trade-off between the biomass (x-axis, h$^{-1}$) and the synthetic objective (y-axis, mmol h$^{-1}$ gDW$^{-1}$). The area underlying each of the ten fronts of non-dominated points represents the number of metabolic configurations in the solution space that are supported when the objectives are maximised (or minimised).}
\label{fig:pareto_front}
\end{figure}

\section{Perspective: integration with multi-view machine learning approaches}
\label{subsec:4}
Multi-view learning can be described as a sub-division of machine learning methods that aims to merge different aspects of a common problem in a single setting. It is based on principles of maximising the consensus between different viewpoints whilst offsetting the limitations of each view through complementation with the other views {\cite{sun2013survey}}. It is evident that this approach is highly applicable to the context of poly-omic data integration in genome scale metabolic models, owing to the interdependencies and correlations between all types of poly-omic data. At the very least, genomics, transcriptomics and proteomics are inextricably linked by the central dogma of molecular biology {\cite{frolova2016integrative}}.

However, as omic types significantly differ in their scale and structure, the data is classed as heterogeneous and several normalisation measures would be required prior to mapping each omic as a layer in the metabolic model. Integration following the simultaneous analysis of multiple data types (known as meta-dimensional analysis) may be preceded by directly concatenating single sample matrices into one large matrix, transforming samples into intermediate graph or kernel matrices, or generating multiple models using different data types as training sets {\cite{ritchie2015methods}}. 

Data integration can be performed at an early, intermediate or late stage, depending on the nature of the data and the learning algorithm used. Early integration allows for the creation of a large pool of data before processing. Intermediate integration involves condensing each view into a similarity matrix with pairwise comparisons before a learning algorithm is applied. Late integration provides the opportunity to select the most suitable learning algorithm for each omic before merging data, comparing analyses between omics and linking patterns found in each omic {\cite{serra2015mvda}}. As early integration increases data dimensionality rather than reducing it, intermediate or late integration would be preferred in this context to avoid introducing noise and decreasing performance prior to data transformation. 

Methods such as clustering and multiple kernel learning have successfully been implemented in cancer sub-typing on the basis of shared molecular characteristics, and are ideal for classifying unsupervised data into groups to detect underlying associations where there is little information available {\cite{taskesen2016pan,speicher2015integrating}}. $K$-means is a traditional clustering algorithm that finds the number of clusters minimising the sum of the squared Euclidean distances between each observation and its respective cluster mean {\cite{mclachlan2008clustering}}. The algorithm starts by selecting $k$ random points in the dataset (termed cluster centroids), which define the groups that the remaining data points are assigned to. The centroids are then moved to the averages computed in each group, and the process is repeated until distinctive clusters are formed. A number of flaws can be identified when this process is applied to multi-view learning. The primary concern is that of a lack of consideration of the importance of each individual view, as well as the differences between multiple views. 

To address these issues, a bi-level variant of $k$-means clustering known as Tw-$k$-means was established, which added simultaneous weighting of both views and individual variables. This resulted in easier identification of the importance of variables and views, as well as in a decrease in the effect of low quality views and noise {\cite{chen2013automated}}. An alternative multi-view clustering approach known as iCluster {\cite{shen2009integrative}} has been developed to model cancer subtypes from the Cancer Genome Atlas. In iCluster, cancer subtypes are considered as latent variables, which could be estimated by taking differences between views into account when partitioning multidimensional data into disparate groups. If a clustering algorithm is chosen for the partitioning of poly-omic data, it is important to note that the clusters of genes or metabolites may vary depending on the condition modelled. Integration by matrix factorisation (IMF) compiles clusters into matrices, which are factorised to assess the contribution of each separate cluster to each view, as well as the overall contribution of each view {\cite{greene2009matrix}}. 

Multiple kernel learning transforms data structures into kernel matrices and optimises weight vectors that linearly combine these matrices to generate a unified kernel matrix. This facilitates the intermediate integration of data from different views, irrespective of the number of features utilised {\cite{speicher2015integrating}}. 
Another kernel-based technique known as similarity network fusion (SNF) {\cite{wang2014similarity}} separately combines samples within each type of data to form individual networks. Such networks are iteratively integrated into a large, comprehensive network, mapped to the feature space in a non-linear fashion and used to assess the amount of information of each data type in explaining any similarity observed between the samples. In a modified version of similarity network fusion {\cite{angione2016multiplex}}, a bias layer was introduced between omic layers to account for the varying quality of metabolic reconstructions and therefore assign larger weights to omic layers that contributed more to the phenotype. 

A support vector machine (SVM) is a supervised learning technique that, given a set of training examples, determines the optimal hyperplane to separate classes in the feature space whilst maximising the distance between samples of different classes {\cite{wasito2012kernel}}. Linear models such as SVMs or the least absolute shrinkage and selection operator (LASSO) can be treated as classification or regression problems respectively for performing feature concatenation with heterogeneous features, such as those found in poly-omic data {\cite{li2016machine}}. 

Decision tree-based methods are highly intuitive and allow the analysis of both continuous and discrete features in the same model without the need for data normalisation. However, the high dimensionality of poly-omic datasets would make decision trees prone to noise and overfitting (especially if there are an insufficient number of features) {\cite{li2016machine}}. This can be overcome by utilising ensemble learning methods such as random forest, which selects features at random as it constructs a decision tree. If a classification approach is taken, the most popular class is voted for following the generation of multiple trees; if the regression approach is taken, outputs from the multiple trees are averaged {\cite{breiman2001random}}.

Feature selection utilises the minimum information necessary to classify key features and could prove useful in identifying the most significant trends in poly-omic datasets. Fortino et al.~{\cite{fortino2014robust}} have described a multivariate discovery process using ``fuzzy patterns'' to discretise and label gene expression data for the selection of the most relevant features. A random forest algorithm was then used to rank features in order of their usefulness, and to improve the stability and accuracy of data. A variation of this method has been used to implement feature selection in the integration of metabolomic, lipidomic and clinical data for the study of obesity and metabolic syndrome {\cite{acharjee2016integration}}.
 
A number of Bayesian methods could also be applied to poly-omic data integration, such as Gaussian mixture models and Latent Dirichlet Allocation (LDA). Angione et al.~{\cite{angione2015hybrid}} incorporated matrix factorisation into a Bayesian hierarchical model using Gaussian Markov Random Fields (GMRFs). This approach led to inferring cross-correlations between pathways in a metabolic network, and to prediction of pathway activation profiles as a result of bacterial responses to environmental conditions. Using a Bayesian method, predefined reaction-pathway memberships were used to model the prior distribution that was multiplied by the likelihood of observations to obtain the posterior distribution, which was subsequently used to infer model predictions. This approach serves to facilitate the observation of links between reactions, pathways and conditions, which can in turn help to interpret the consequences of metabolic flux variations and, consequently, the biological system as a whole. LDA has yet to be applied to poly-omic integration but has potential because of its efficacy in organising unsupervised data from a mixture of clusters {\cite{pratanwanich2014exploring}}.

\section{Conclusion}
\label{sec:conclusion}

In this work, we have surveyed the principal methods available for constraint-based modelling and omic integration. We have presented these in the form of a `forest', also available in interactive version at \url{http://modellingmetabolism.net} where it will be updated periodically. As an up-to-date classification of available methods, we believe that this will prove to be a useful resource for prospective modellers. We have also provided the first tutorial in R for multi-objective optimisation of metabolic models.

We envisage that a late integration approach can be used to test the suitability of various multi-view learning algorithms for each omic dataset used in the integration process, depending on the structure of the data. For example, a clustering algorithm may be chosen for mapping microarray or RNA sequencing data onto a metabolic model. Likewise, a SVM may be chosen for mapping protein data or post-translational modifications. Finally, feature selection may be employed to identify the most distinctive features across both of these omic layers, as a whole. 

In the near future, the integration of multi-view machine learning in metabolic modelling is likely to grow in parallel with the rapid advancement of high-throughput omic technologies. With the increase in the number of omic layers, existing algorithms will continue to be extended. 
However, we believe a one-size-fits-all approach is not the most effective way of tackling a multi-omic genome-scale problem. Given the intrinsic differences between omic layers, we envisage specific algorithms designed only for the analysis of each layer. When required, a global prediction can then be achieved through methods for aggregation of layers, such as those successfully employed in multilayer network theory.

\bibliographystyle{vancouver_briefings}
\bibliography{refs}

\appendix

\section{Full code and details of the R tutorial on genetic design by multi-objective optimisation}

Here we present and describe the full code needed to conduct Genetic Design by Multi-Objective Optimisation in R. Note that here, in addition to the functionality described in the main text, we here cover the boilerplate code associated with loading libraries and preparing data, and the more length aspects of the NSGA-II procedure.

First, we need to load the appropriate libraries:

\begin{itemize}
\itemsep1pt\parskip0pt\parsep0pt
\item
  \texttt{tidyverse} is a bundle of generic utilities;
\item
  \texttt{stringr} is a string manipulation utility installed alongside
  \texttt{tidyverse};
\item
  \texttt{fbar} is a library for flux balance analysis;
\end{itemize}

\begin{Shaded}
\begin{Highlighting}[]
\KeywordTok{library}\NormalTok{(tidyverse)}
\KeywordTok{library}\NormalTok{(stringr)}
\KeywordTok{library}\NormalTok{(fbar)}
\end{Highlighting}
\end{Shaded}

This code block downloads and reads in a model, then extracts the list
of genes from the model. The model takes the form of a tabular list of
reactions.

\begin{Shaded}
\begin{Highlighting}[]
\NormalTok{model <-}\StringTok{ }\KeywordTok{read_tsv}\NormalTok{(}\StringTok{'https://git.io/v1YsM'}\NormalTok{, }
                  \DataTypeTok{col_types =} \KeywordTok{c}\NormalTok{(}\StringTok{'cccdddc'}\NormalTok{))}

\NormalTok{genes_in_model <-}\StringTok{ }\NormalTok{model$geneAssociation %>%}
\StringTok{  }\KeywordTok{str_split}\NormalTok{(}\StringTok{'[()|& ]+'}\NormalTok{) %>%}
\StringTok{  }\KeywordTok{flatten_chr}\NormalTok{() %>%}
\StringTok{  }\KeywordTok{discard}\NormalTok{(is.na) %>%}
\StringTok{  }\KeywordTok{discard}\NormalTok{(~}\StringTok{ }\KeywordTok{str_length}\NormalTok{(.x)==}\DecValTok{0}\NormalTok{)}
\end{Highlighting}
\end{Shaded}

The evaluation function is where the actual metabolic simulations are
performed. This has four main stages:
\begin{enumerate}
\itemsep1pt\parskip0pt\parsep0pt
\item
  The gene-reaction associations (\texttt{geneAssociation}) are
  evaluated in the context of which genes are present in this iteration
  (\texttt{genome}), to give an \texttt{activation} value, which is an estimate of reaction rate.
\item
  The \texttt{activation} value is used to alter the upper and lower bounds on reaction rate (\texttt{uppbnd} and \texttt{lowbnd}), to push reaction rates towards the rate estimates.
\item
  We conduct a round of FBA, optimising for maximum biomass.
\item
  We fix the biomass production value to its maximum by altering the corresponding \texttt{uppbnd} and \texttt{lowbnd} to be near the \texttt{flux} (+/-1\%).
\item
  With the biomass value fixed, we alter the objective coefficient (\texttt{obj\_coef}) to target optimisation of the synthetic objective.
\end{enumerate}

The technique of fixing the biomass followed by maximising the synthetic
objective is important because there could still be slack in the model
after the first optimisation stage, and we wish to have a reliable
estimate of the synthetic objective.

\begin{Shaded}
\begin{Highlighting}[]
\NormalTok{evaluation_function <-}\StringTok{ }\NormalTok{function(genome)\{}
  
  \NormalTok{res <-}\StringTok{ }\NormalTok{model %>%}
\StringTok{    }\KeywordTok{mutate}\NormalTok{(}\DataTypeTok{activation =} \NormalTok{fbar::}\KeywordTok{gene_eval}\NormalTok{(}\DataTypeTok{expressions =} \NormalTok{geneAssociation, }
                                        \DataTypeTok{genes =} \KeywordTok{names}\NormalTok{(genome), }
                                        \DataTypeTok{presences =} \NormalTok{genome}
                                        \NormalTok{),}
           \DataTypeTok{activation =} \KeywordTok{coalesce}\NormalTok{(activation, }\DecValTok{1}\NormalTok{),}
           \DataTypeTok{uppbnd =} \KeywordTok{pmin}\NormalTok{(uppbnd, }\DecValTok{1000}\NormalTok{*activation}\FloatTok{+0.1}\NormalTok{),}
           \DataTypeTok{lowbnd =} \KeywordTok{pmax}\NormalTok{(lowbnd, -}\DecValTok{1000}\NormalTok{*activation}\FloatTok{-0.1}\NormalTok{)) %>%}
\StringTok{    }\NormalTok{fbar::}\KeywordTok{find_fluxes_df}\NormalTok{(}\DataTypeTok{do_minimization =} \OtherTok{FALSE}\NormalTok{) %>%}
\StringTok{    }\KeywordTok{mutate}\NormalTok{(}\DataTypeTok{lowbnd =} \KeywordTok{ifelse}\NormalTok{(abbreviation==}\StringTok{'Biomass_Ecoli_core_w/GAM'}\NormalTok{, }
                           \NormalTok{flux*}\FloatTok{0.99}\NormalTok{, }
                           \NormalTok{lowbnd),}
           \DataTypeTok{uppbnd =} \KeywordTok{ifelse}\NormalTok{(abbreviation==}\StringTok{'Biomass_Ecoli_core_w/GAM'}\NormalTok{, }
                           \NormalTok{flux*}\FloatTok{1.01}\NormalTok{, }
                           \NormalTok{uppbnd),}
           \DataTypeTok{obj_coef =} \DecValTok{1}\NormalTok{*(abbreviation==}\StringTok{'EX_ac(e)'}\NormalTok{)) %>%}
\StringTok{    }\NormalTok{fbar::}\KeywordTok{find_fluxes_df}\NormalTok{(}\DataTypeTok{do_minimization =} \OtherTok{FALSE}\NormalTok{)}
  
  \KeywordTok{return}\NormalTok{(}\KeywordTok{list}\NormalTok{(}\DataTypeTok{bm =} \KeywordTok{filter}\NormalTok{(res, abbreviation==}\StringTok{'Biomass_Ecoli_core_w/GAM'}\NormalTok{)$flux, }
              \DataTypeTok{synth =} \KeywordTok{filter}\NormalTok{(res, abbreviation==}\StringTok{'EX_ac(e)'}\NormalTok{)$flux))}
\NormalTok{\}}
\end{Highlighting}
\end{Shaded}

Non-domination sorting is the first stage of the selection procedure in
NSGA-II. The code might seem quite opaque, but the idea is as follows:
\begin{enumerate}
\itemsep1pt\parskip0pt\parsep0pt
\item
  We perform an \texttt{inner\_join} in order to compare every point against
  every other point.
\item
  For each point (\texttt{id.x}), we see if there exists any second
  point (\texttt{id.y}) that has a higher value in all
  objectives. Where such a second point exists, we term the original
  point `dominated'.
\item
  We find the set of points that have no dominating point, and term
  this the first non-dominated front.
\item
  We repeat this procedure, but ignore points in the first
  non-dominated front to find the second non-dominated front, and so on.
\end{enumerate}

\begin{Shaded}
\begin{Highlighting}[]
\NormalTok{non_dom_sort <-}\StringTok{ }\NormalTok{function(input)\{}
  \NormalTok{input_long <-}\StringTok{ }\NormalTok{input %>%}
\StringTok{    }\KeywordTok{gather}\NormalTok{(property, value, -id) %>%}
\StringTok{    }\KeywordTok{mutate}\NormalTok{(}\DataTypeTok{front=}\OtherTok{NA}\NormalTok{)}
  
  \NormalTok{currentfront <-}\StringTok{ }\DecValTok{1}
  
  \NormalTok{while(}\KeywordTok{any}\NormalTok{(}\KeywordTok{is.na}\NormalTok{(input_long$front)))\{}
    
    \NormalTok{input_long <-}\StringTok{ }\NormalTok{input_long %>%}
\StringTok{      }\KeywordTok{inner_join}\NormalTok{(.,., }\DataTypeTok{by=}\StringTok{'property'}\NormalTok{) %>%}
\StringTok{      }\KeywordTok{group_by}\NormalTok{(id.x,id.y) %>%}
\StringTok{      }\KeywordTok{mutate}\NormalTok{(}\DataTypeTok{dominance =} \KeywordTok{ifelse}\NormalTok{(}\KeywordTok{all}\NormalTok{(value.x>=value.y), }
                                \StringTok{'xdomy'}\NormalTok{, }
                                \KeywordTok{ifelse}\NormalTok{(}\KeywordTok{all}\NormalTok{(value.y>=value.x), }
                                       \StringTok{'ydomx'}\NormalTok{, }
                                       \StringTok{'nondom'}
                                       \NormalTok{)}
                                \NormalTok{)}
      \NormalTok{) %>%}
\StringTok{      }\KeywordTok{group_by}\NormalTok{(id.x) %>%}
\StringTok{      }\KeywordTok{mutate}\NormalTok{(}\DataTypeTok{front =} \KeywordTok{ifelse}\NormalTok{(}\KeywordTok{all}\NormalTok{(dominance[}\KeywordTok{is.na}\NormalTok{(front.y)] %in%}\StringTok{ }\KeywordTok{c}\NormalTok{(}\StringTok{'xdomy'}\NormalTok{, }\StringTok{'nondom'}\NormalTok{)), }
                            \KeywordTok{pmin}\NormalTok{(currentfront, front.x, }\DataTypeTok{na.rm=}\OtherTok{TRUE}\NormalTok{), }
                            \OtherTok{NA}
                            \NormalTok{)}
      \NormalTok{) %>%}
\StringTok{      }\KeywordTok{group_by}\NormalTok{(}\DataTypeTok{id =} \NormalTok{id.x, }\DataTypeTok{property =} \NormalTok{property, front, }\DataTypeTok{value =} \NormalTok{value.x) %>%}
\StringTok{      }\NormalTok{summarise}
    
    \NormalTok{currentfront <-}\StringTok{ }\NormalTok{currentfront +}\StringTok{ }\DecValTok{1}
  \NormalTok{\}}
  
  \KeywordTok{return}\NormalTok{(}
    \NormalTok{input_long %>%}
\StringTok{      }\KeywordTok{spread}\NormalTok{(property, value)}
  \NormalTok{)}
  
\NormalTok{\}}
\end{Highlighting}
\end{Shaded}

The second part of the NSGA-II evaluation procedure is finding the
crowding distance. This is used to break ties between points in the same
non-dominated front. For each front and for each dimension, this
function sorts the points into order along the dimension, and finds the
normalised distance between the proceeding point and succeeding point.
These values are summed up across each dimension to find the value for
the point.

\begin{Shaded}
\begin{Highlighting}[]
\NormalTok{crowding_distance <-}\StringTok{ }\NormalTok{function(input)\{}
  \KeywordTok{return}\NormalTok{(}
    \NormalTok{input %>%}
\StringTok{      }\KeywordTok{gather}\NormalTok{(property, value, -id, -front) %>%}
\StringTok{      }\KeywordTok{group_by}\NormalTok{(front, property) %>%}
\StringTok{      }\KeywordTok{arrange}\NormalTok{(value) %>%}
\StringTok{      }\KeywordTok{mutate}\NormalTok{(}\DataTypeTok{crowding =} \NormalTok{(}\KeywordTok{lead}\NormalTok{(value)-}\KeywordTok{lag}\NormalTok{(value))/(}\KeywordTok{max}\NormalTok{(value)-}\KeywordTok{min}\NormalTok{(value)),}
             \DataTypeTok{crowding =} \KeywordTok{ifelse}\NormalTok{(}\KeywordTok{is.na}\NormalTok{(crowding),}\OtherTok{Inf}\NormalTok{, crowding)) %>%}
\StringTok{      }\KeywordTok{group_by}\NormalTok{(id) %>%}
\StringTok{      }\KeywordTok{mutate}\NormalTok{(}\DataTypeTok{crowding =} \KeywordTok{sum}\NormalTok{(crowding)) %>%}
\StringTok{      }\KeywordTok{spread}\NormalTok{(property, value)}
  \NormalTok{)}
\NormalTok{\}}
\end{Highlighting}
\end{Shaded}

The following code is the genetic loop of the algorithm. It is explained by code comments, but follows a normal pattern of evaluating, sorting, selecting from and mutating the population. The genetic algorithm used here is a modified version of NSGA-II \cite{deb2002fast}, with a population of $200$ individuals and carrying out $500$ iterations.

\begin{Shaded}
\begin{Highlighting}[]
\NormalTok{start_genome <-}\StringTok{ }\KeywordTok{set_names}\NormalTok{(}\KeywordTok{rep_along}\NormalTok{(genes_in_model, }\OtherTok{TRUE}\NormalTok{), genes_in_model)}
\NormalTok{pop <-}\StringTok{ }\KeywordTok{list}\NormalTok{(start_genome)}

\NormalTok{popsize =}\StringTok{ }\DecValTok{200}
\NormalTok{generations =}\StringTok{ }\DecValTok{500}

\NormalTok{pb <-}\StringTok{ }\KeywordTok{txtProgressBar}\NormalTok{(}\DataTypeTok{max=}\NormalTok{generations, }\DataTypeTok{style=}\DecValTok{3}\NormalTok{)}
\NormalTok{for(i in }\DecValTok{1}\NormalTok{:generations)\{}
  \KeywordTok{setTxtProgressBar}\NormalTok{(pb, i)}
  \NormalTok{results <-}\StringTok{ }\KeywordTok{map_df}\NormalTok{(pop, evaluation_function) %>%}\StringTok{ }\CommentTok{# Evaluate all the genomes}
\StringTok{    }\KeywordTok{mutate}\NormalTok{(}\DataTypeTok{bm=}\KeywordTok{signif}\NormalTok{(bm), }\DataTypeTok{synth=}\KeywordTok{signif}\NormalTok{(synth)) %>%}\StringTok{ }\CommentTok{# Round results}
\StringTok{    }\KeywordTok{mutate}\NormalTok{(}\DataTypeTok{id =} \DecValTok{1}\NormalTok{:}\KeywordTok{n}\NormalTok{()) %>%}\StringTok{ }\CommentTok{# label the results}
\StringTok{    }\KeywordTok{sample_frac}\NormalTok{() %>%}\StringTok{ }\CommentTok{# Shuffle}
\StringTok{    }\KeywordTok{non_dom_sort}\NormalTok{() %>%}\StringTok{ }\CommentTok{# Find the non-dominated fronts}
\StringTok{    }\KeywordTok{crowding_distance}\NormalTok{() %>%}\StringTok{ }\CommentTok{# Find the crowding distances}
\StringTok{    }\KeywordTok{arrange}\NormalTok{(front, }\KeywordTok{desc}\NormalTok{(crowding)) }\CommentTok{# Sort by front, breaking ties by crowding distance}
  
  \NormalTok{selected <-}\StringTok{ }\NormalTok{results %>%}
\StringTok{    }\KeywordTok{filter}\NormalTok{(}\KeywordTok{row_number}\NormalTok{() <=}\StringTok{ }\NormalTok{popsize/}\DecValTok{2}\NormalTok{) %>%}\StringTok{ }\CommentTok{# Keep the best half of the population}
\StringTok{    }\KeywordTok{getElement}\NormalTok{(}\StringTok{'id'}\NormalTok{)}
  
  \NormalTok{kept_pop <-}\StringTok{ }\NormalTok{pop[selected]}
  
  \NormalTok{altered_pop <-}\StringTok{ }\NormalTok{kept_pop %>%}
\StringTok{    }\KeywordTok{sample}\NormalTok{(popsize-}\KeywordTok{length}\NormalTok{(selected), }\OtherTok{TRUE}\NormalTok{) %>%}\StringTok{ }\CommentTok{# Sample parents from population}
\StringTok{    }\KeywordTok{map}\NormalTok{(function(genome)\{}
      \KeywordTok{xor}\NormalTok{(genome, }\KeywordTok{runif}\NormalTok{(}\KeywordTok{length}\NormalTok{(genome))>}\FloatTok{0.98}\NormalTok{) }\CommentTok{# Mutate parents to create offspring}
      \NormalTok{\})}
  
  \NormalTok{pop <-}\StringTok{ }\KeywordTok{unique}\NormalTok{(}\KeywordTok{c}\NormalTok{(kept_pop, altered_pop)) }\CommentTok{# Combine the offspring and parent populations}
\NormalTok{\}}
\end{Highlighting}
\end{Shaded}


Once we have a results set, we can construct a plot to view the
non-dominated fronts. We can see how the first front describes the
trade-off between biomass and the synthetic objective, with the lines
showing the dominated area (to the bottom left). 

\begin{Shaded}
\begin{Highlighting}[]
\NormalTok{pop %>%}
\StringTok{  }\KeywordTok{arrange}\NormalTok{(}\KeywordTok{desc}\NormalTok{(front)) %>%}
\StringTok{  }\KeywordTok{ggplot}\NormalTok{(}\KeywordTok{aes}\NormalTok{(}\DataTypeTok{x=}\NormalTok{bm, }\DataTypeTok{y=}\NormalTok{synth, }\DataTypeTok{colour=}\KeywordTok{factor}\NormalTok{(front))) +}\StringTok{ }
\StringTok{  }\KeywordTok{geom_point}\NormalTok{() +}\StringTok{ }\KeywordTok{geom_step}\NormalTok{(}\DataTypeTok{direction=}\StringTok{'vh'}\NormalTok{, }\DataTypeTok{alpha=}\FloatTok{0.5}\NormalTok{) +}
\StringTok{  }\KeywordTok{theme_bw}\NormalTok{()}
\end{Highlighting}
\end{Shaded}

\end{document}